\documentclass[12pt]{article}
\usepackage{amsmath,amssymb,geometry,microtype,amsthm}
\usepackage{marginnote,xparse,changepage,caption,graphicx,subcaption}
\usepackage[toc,page]{appendix}
\usepackage[backend=biber,sorting=none]{biblatex}
\usepackage{tocloft}

\newtheorem{claim}{Claim}

\newcommand{\mC}{\mathcal{C}}
\newcommand{\mH}{\mathcal{H}}
\newcommand{\mM}{\mathcal{M}}
\newcommand{\mJ}{\mathcal{J}}
\newcommand{\IGNORE}[1]{}
\newcommand{\be}{\begin{equation}}
\newcommand{\ee}{\end{equation}}
\newcommand{\bea}{\begin{aligned}}
\newcommand{\eea}{\end{aligned}}
\newcommand{\F}[1]{F_{#1}}
\renewcommand{\P}[1]{P_{#1}}

\usepackage{todonotes}
\usepackage{soul}

\RequirePackage[colorlinks=true
,urlcolor=blue
,anchorcolor=blue
,citecolor=blue
,filecolor=blue
,linkcolor=blue
,menucolor=blue
,pagecolor=blue
,linktocpage=true
,pdfproducer=medialab
]{hyperref}

\graphicspath{ {./} }

\numberwithin{equation}{section}

\begin{document}

\vfill

\begin{center}
   \baselineskip=16pt {\Large\bf Horizon Molecules in Causal Set Theory}
  \vskip 1.5cm {Christopher Barton${}^a$, Andrew Counsell${}^a$,
      Fay Dowker${}^{a,b}$, Dewi S. W. Gould${}^{a}$, Ian~Jubb${}^{c}$, and Gwylim Taylor${}^{a}$\\
   \vskip .6cm}
  \begin{small}
      \textit{
      		${}^{a}${Blackett Laboratory, Imperial College, Prince Consort Road, London, SW7 2AZ}\\\vspace{5pt}
		${}^{b}${Perimeter Institute, 31 Caroline Street North, Waterloo ON, N2L 2Y5, Canada}\\\vspace{5pt}
		${}^{c}${School of Theoretical Physics, Dublin Institute for Advanced Studies, 10 Burlington Road, D04 C932, Ireland}
              }
              \end{small}\\*[.6cm]
   \end{center}

\vspace{20pt}

\begin{center}
\textbf{Abstract}
\end{center}

\begin{quote}
We propose a new definition of ``horizon molecules" in Causal Set Theory following pioneering work by Dou and Sorkin. The new concept applies for any causal horizon and its intersection with any spacelike hypersurface.  In the continuum limit, as the discreteness scale tends to zero, the leading behaviour of the expected number of horizon molecules is shown to be the area of the horizon in discreteness units, up to a dimension dependent factor of order one.  We also determine the first order corrections to the continuum value, and show how such corrections can be exploited to obtain further geometrical information about the horizon and the spacelike hypersurface from the causal set. 
\end{quote}

\pagebreak
\tableofcontents

\newpage
\section{Introduction}\label{sec:intro}

The idea of counting ``horizon molecules'' in a causal set (causet) approximated by a black hole spacetime, in order to estimate the black hole entropy,
was pioneered by Dou and Sorkin (DS) \cite{Dou:2003af}~\footnote{See~\cite{Surya:2019ndm} for an up-to-date review of Causal Set Theory.}. According to DS:  `` [T]he picture of the horizon as
composed of discrete constituents gives a good account of the entropy if we
suppose that each such constituent occupies roughly one unit of Planck area
and carries roughly one bit of entropy. A proper statistical derivation along
these lines would require a knowledge of the dynamics of these constituents,
of course. However, in analogy with [a] gas, one may still anticipate that the
horizon entropy can be estimated by counting suitable discrete structures,
analogues of the gas molecules, without referring directly to their dynamics.''

The original proposal of DS was that a horizon molecule should be the 
simplest possible subcauset that is not a single causet element, namely a causal link.
A link is a subcauset of cardinality 2 in which the 2 elements are 
related, and such that no other element of the causet is 
between them in the order. DS proposed that the lower (minimal) element of the link should be outside the horizon and the upper (maximal) element should be inside in order to do justice to the idea of the black hole entropy arising, at least partly if not wholly, from entanglement between degrees of freedom inside and outside the horizon \cite{Sorkin:1985bu}.

The DS proposal gave promising answers in the case of 2-dimensional truncations of a Schwarzschild black hole and of the dynamical horizon of a spherically symmetric collapsing shell. Both cases gave the same leading constant term for the expected value of the number of molecules. However,
it was realised by Dou \cite{Dou:priv} that the proposed molecules would not work in higher dimensions: in 3 or more dimensions the number of DS horizon molecules 
is unbounded for a black hole in an infinite environment, even at non-zero discreteness scale (this divergence is explained in \cite{Marr:2007}.) This led to a number of new proposals for horizon molecules of cardinality 3 and 4 \cite{Marr:2007, students}. These new proposals did not suffer the same divergence in higher dimensions as the DS molecule but because of the more complicated molecule structure, the calculations involved in determining the expected number in greater than 2 dimensions are challenging. 
Proof is still lacking that counting one of these higher cardinality molecules gives the horizon area as desired in greater than 2 spacetime dimensions. 

One feature of the DS proposal is that the definition of horizon molecule is the same whether the hypersurface $\Sigma$ on which the entropy of the black hole is evaluated is spacelike \textit{or} null. Indeed the successful calculations of DS in 1+1 dimensions were actually done for  null $\Sigma$.  The higher cardinality molecule definitions of Marr and others
were also for $\Sigma$ null or spacelike.  The calculational impasse, a desire to extend the concept of horizon molecule to all \textit{causal horizons} including black hole, acceleration, and cosmological horizons \cite{Gibbons:1977, Jacobson:2003wv}, and a desire to return to the original attractive DS conception of a molecule as a simple \textit{link} straddling the horizon, stimulated a fresh look at the problem. The key to the progress reported in the current paper was to require the definition of horizon molecule to work when the hypersurface $\Sigma$ is spacelike, but not to require it to work when $\Sigma$ is null. 

\section{The proposal}\label{sec:Setup and definitions}

Let  $ (\mathcal{M},g)$ be a $d$-dimensional globally hyperbolic spacetime with a Cauchy surface $\Sigma$ which intersects a {causal horizon} $\mathcal{H}$ in a co-dimension 2 spacelike surface $\mathcal{J} = \mathcal{H} \cap \Sigma$. In a nod to the importance of the $d=4$ case we will refer to the $(d-2)$-volume of the intersection 
$\mathcal{J}$ as the \textit{area} of $\mathcal{J}$.

$\mathcal{H}$ is a causal horizon. \textit{i.e.} it is the boundary of the past of a future inextendible timelike curve, $\gamma_0$, of infinite proper future length: $\mH := \partial I^-(\gamma_0)$.  To $\mH$ we can associate a past set, $\mM_-$, and a future set, $\mM_+$, which, together with $\mH$, partition $\mM$: 
\begin{align*}
 \mM_- &: = I^-(\gamma_0)\,,\\
 \mM_+ &: = \mM \setminus (\mM_- \cup \mH)\,.
 \end{align*}
 
\begin{figure}[t]
\caption{\footnotesize{An illustration of the geometric setup in $d=3$.}}
\includegraphics[trim={0 4cm 0 4cm},clip,width=12cm]{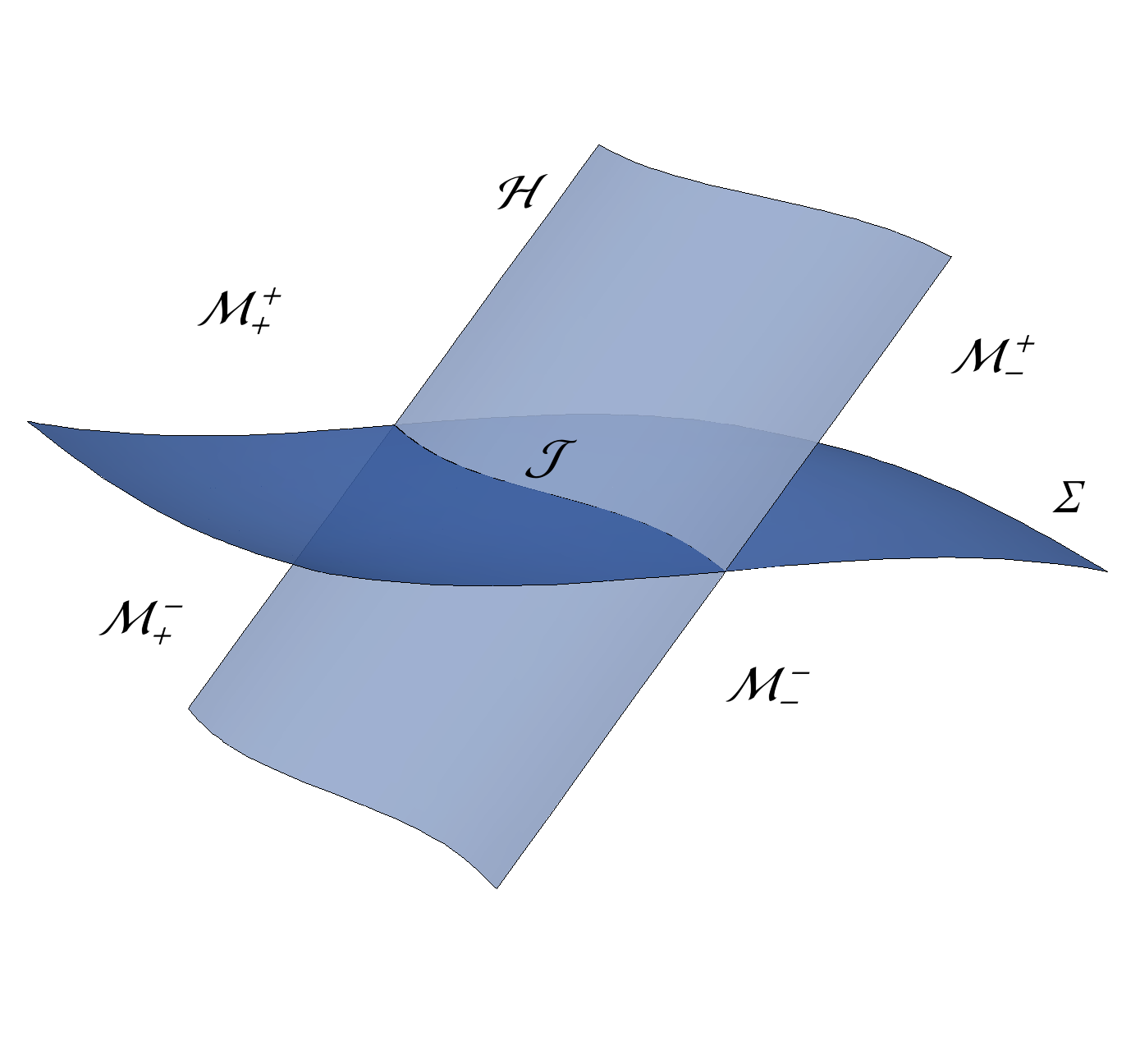}\label{fig:geometric_setup}
\centering
\end{figure}

 \noindent To $\Sigma$ we associate past and future sets, $\mM^-$ and $\mM^+$ respectively: $\mM^{\pm} : = I^\pm(\Sigma)$. Again these two sets, together with $\Sigma$, partition $\mM$. If we take intersections of these partitions we obtain the 4 regions $\mM_{\pm}^\pm : = \mM^{\pm} \cap \mM_\pm$ sketched in figure \ref{fig:geometric_setup}.

Following DS, we consider the Poisson point process of sprinkling at density $\rho = l^{-d}$ into $\mathcal{M}$. This process results in a random causet $(\mC, \prec)$ which is a possible substratum to which the continuum $(\mM$,g) is an approximation at scales much larger than the discreteness scale $l$. The subcausets of $\mC$ sprinkled into the regions 
$\mM^\pm$, $\mM_+^-$ \textit{etc.},  are labelled in the obvious way: $\mC^\pm$, 
 $\mC_+^-$ respectively \textit{etc.} We will be interested in the limit $\rho \rightarrow \infty$ (equivalently $l \rightarrow 0$) and the approach to the limit. We 
 will refer to this as the continuum limit. 
 
\begin{figure}[t]
\caption{\footnotesize{An illustration of a horizon molecule. The link between the points $p_{\pm}$ is shown in red. The black dashed lines indicate the future lightcone from $p_-$, from which one can see that only $p_+$ is to its future within the past of $\Sigma$. We have also included another dashed line from $\mathcal{J}$ to illustrate the region corresponding to $I^-(\mathcal{J})$.}}
\includegraphics[width=12cm]{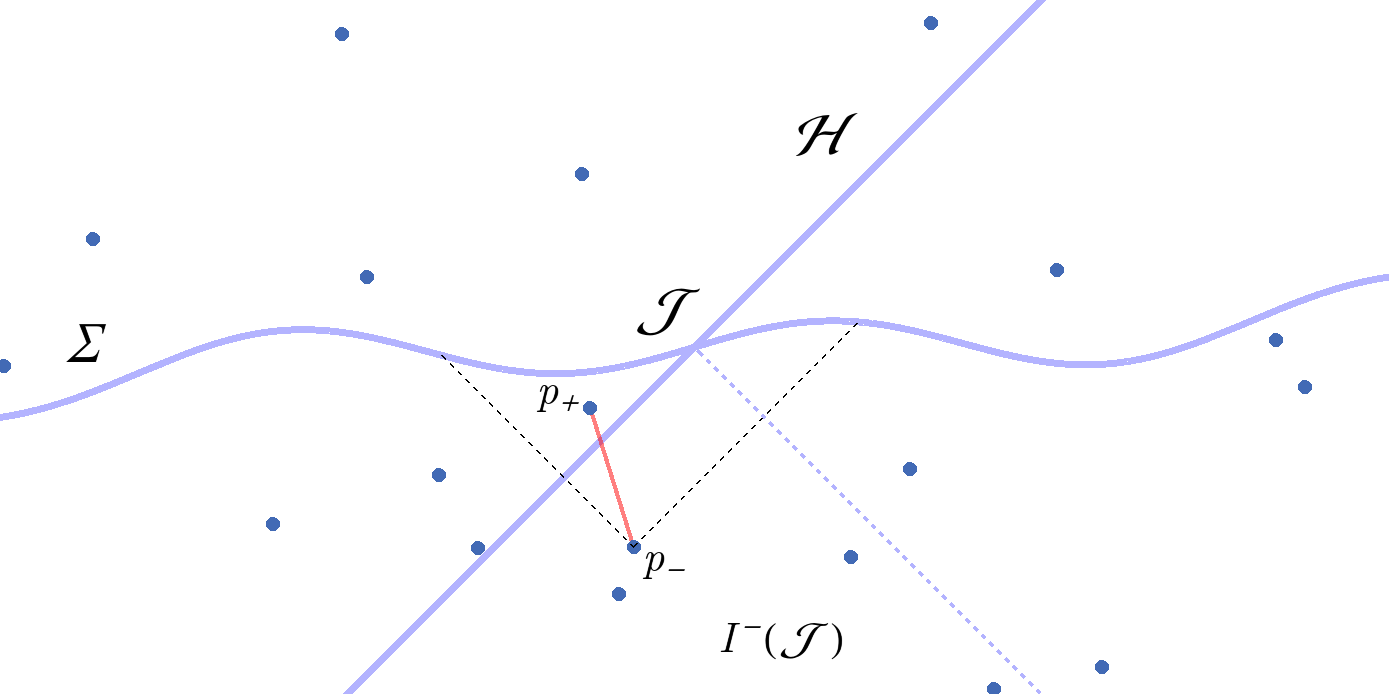}\label{fig:horizon_molecule}
\centering
\end{figure}

 We are interested in the entropy of  $\mH$ on the hypersurface $\Sigma$ and we propose a definition of horizon molecule
for $\mH$,  associated with $\Sigma$, using only the structure
 of $\mC$ and its partitions into $\mC^\pm$ \textit{etc}.:
 \newpage
\newtheorem*{HA}{Definition}
\begin{HA}
A  \emph{ horizon molecule} is a pair of elements of $\mC^-$, $\{p_-, p_+\}$, such that:
\begin{itemize}
\item $p_- \prec p_+$,
\item  $p_-\in\mathcal{C}^-_-$,
\item  $p_+\in\mathcal{C}^-_+$,
\item  $p_+$ is the only element in both $\mathcal{C}^-$ and the future of 
$p_-$. 
\end{itemize}
\end{HA}
\noindent These conditions imply that a horizon molecule is a link. 
See figure~\ref{fig:horizon_molecule} for an illustration of a horizon molecule. More generally, one can define:
 \newtheorem*{HB}{Definition}
\begin{HB} 
A  \emph{horizon} $n$-\emph{molecule} is a subcauset of $\mC^-$, 
$\{p_-, p_{+,1},...,p_{+,n}\}$ such that 
\begin{itemize}
\item $p_- \prec p_{+, k}$ for all $k = 1,2, \dots n$; 
\item  $p_-\in\mathcal{C}^-_-$; 
\item  $p_{+,k} \in\mathcal{C}^-_+$ for all $k = 1,2, \dots n$;
\item  $\{p_{+,1},...,p_{+,n}\} $ are  the only elements in both $\mathcal{C}^-$ and the future of $p_-$. 
\end{itemize}
\end{HB}
\noindent The $1$-molecule  is the molecule defined previously, and seems most natural as a definition of a causal set horizon molecule, but we will give results for $n>1$ also. 

In a given sprinkling, the definition of a horizon $n$-molecule implies that the minimal element $p_-$, of each molecule, lies in the spacetime region $I^-(\mathcal{M}^-_+)\cap \mathcal{M}^-_-$. In appendix~\ref{app:proof_mj_is_past_of_j} we show that this implies $p_-$ is in the chronological past of $\mathcal{J}$, by showing that $I^-(\mathcal{M}^-_+)\cap \mathcal{M}^-_- = I^-(\mathcal{J})$.

For a given causal set embedded in $(\mM, g)$, with $\mH$ and $\Sigma$, we define the number, $\mathrm{H}$, of horizon molecules. Under the sprinkling process, the number $\mathrm{H}$ becomes a random variable, which we denote by $\mathbf{H}$, which depends on the sprinkling density $\rho$, though we don't make that dependence  explicit in the notation. 
For $n$-molecules more generally we define $\mathrm{H}_n$ as the number of $n$-molecules in a sprinkling into $(\mM, g)$. 

We make the following 
\begin{claim}
 In the continuum limit, the expected number of horizon molecules is equal to the 
 area of $\mJ$, the intersection of the horizon and $\Sigma$,  in discreteness units, up to a dimension dependent constant of order one.
\end{claim}

Stated mathematically the claim is
\begin{equation}\label{eq:general_area_result}
\lim_{\rho\rightarrow\infty} \rho^{\frac{2-d}{d}} \langle \mathbf{H} \rangle = a^{(d)}\int_{\mathcal{J}} dV_{\mathcal{J}} \;\;\; ,
\end{equation}
where $\langle \cdot\rangle$ denotes the mean over sprinklings, $dV_{\mathcal{J}}$ is the area measure on $\mathcal{J}$, and $a^{(d)}$ is a constant that only depends on the dimension $d$. 
In the case of infinite causal horizons, such as a Rindler horizon in Minkowski spacetime, (\ref{eq:general_area_result}) is interpreted as saying that there is a fixed, finite, dimension dependent, mean \textit{density} of number of horizon molecules per unit area in discreteness units. 

More generally, for $n$-molecules, we claim 
\begin{equation}\label{eq:general_area_result_n}
\lim_{\rho\rightarrow\infty} \rho^{\frac{2-d}{d}} \langle \mathbf{H}_n \rangle = a^{(d)}_n\int_{\mathcal{J}} dV_{\mathcal{J}} \;\;\; ,
\end{equation}
where $a^{(d)}_n$ depends on $d$ and $n$.

In this paper we prove this result under certain assumptions, argue that the approach to the limit involves finite $\rho$ corrections forming a derivative expansion of local geometric quantities on $\mathcal{J}$ and increasing powers of $l$, the discreteness length.

\subsection{Setting up the calculation}\label{sec:Formulating the causal set expectation value}

\begin{figure}
\caption{\footnotesize{An illustration of the volumes $V(p)$ and $V_+(p)$. We have not shown $V_-(p)$, but this can be worked out from $V_-(p) = V(p) - V_+(p)$.}}
\begin{subfigure}[b]{0.5\textwidth}
\caption{$V(p)$}
\includegraphics[width=7cm]{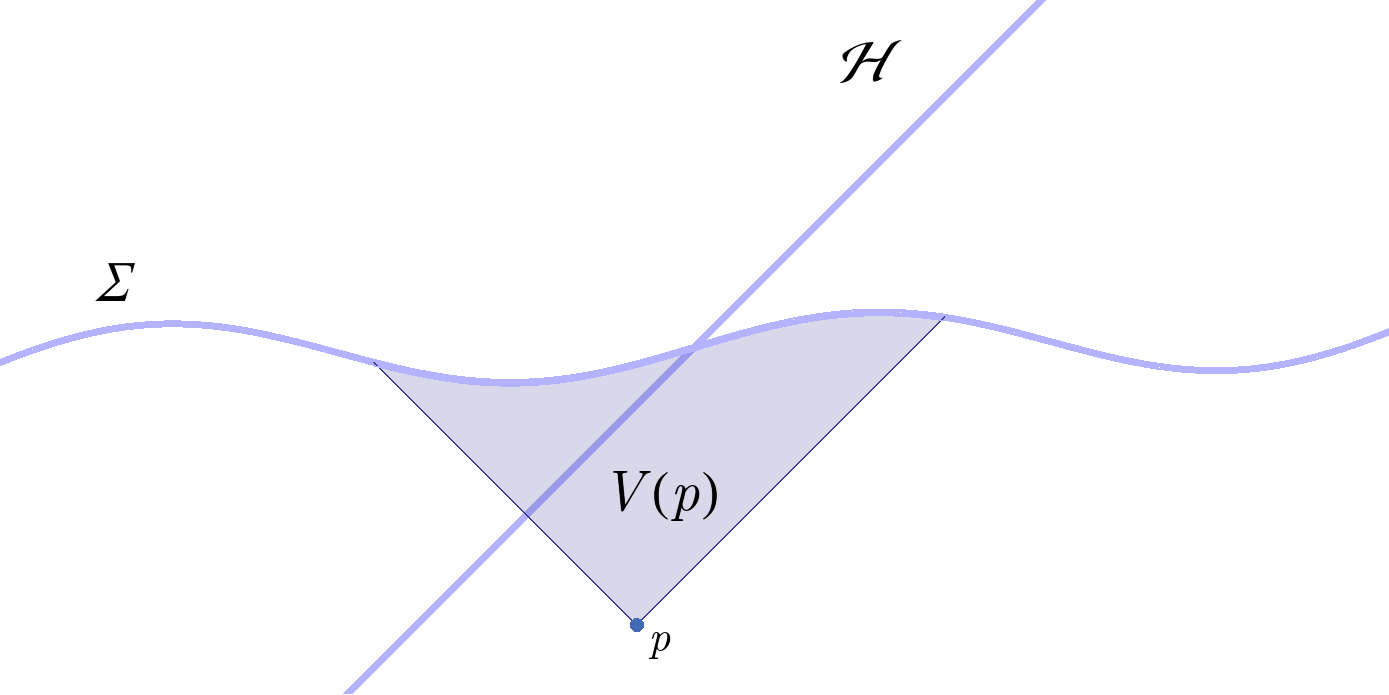}\label{fig:vp}
\centering
\end{subfigure}
\begin{subfigure}[b]{0.5\textwidth}
\caption{$V_+(p)$}
\includegraphics[width=7cm]{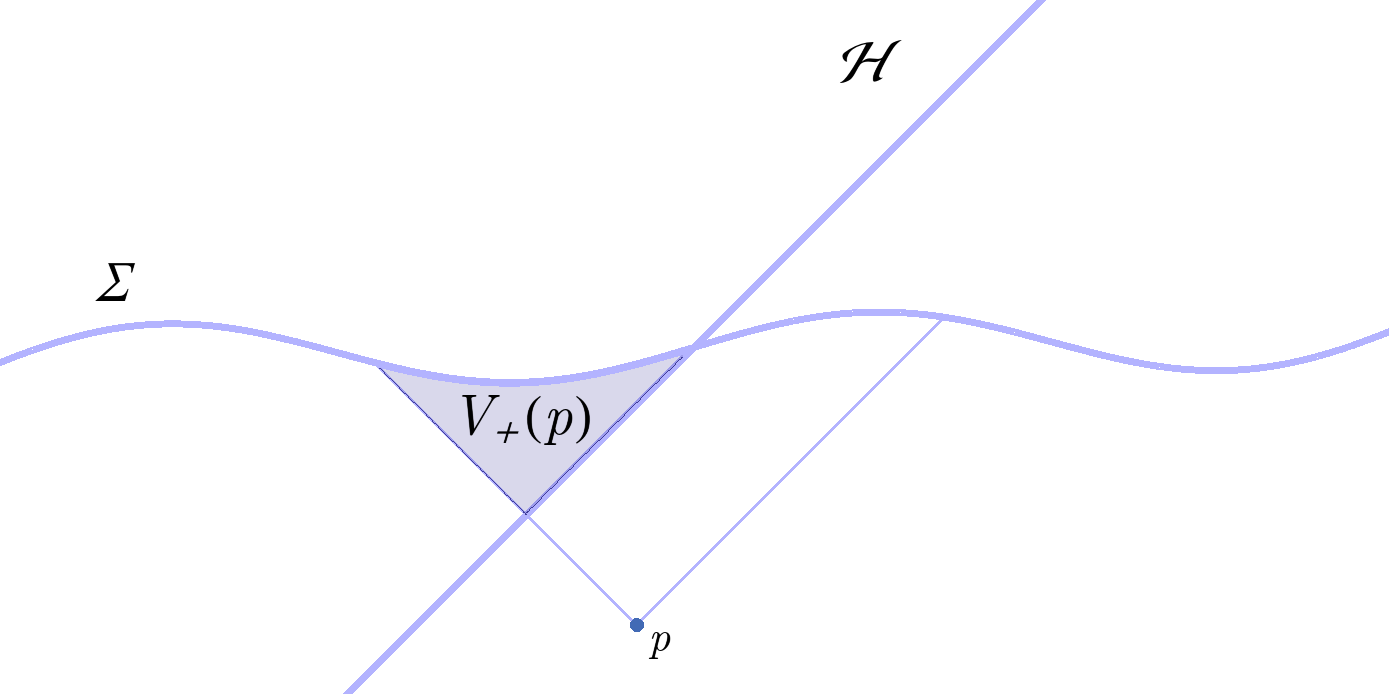}\label{fig:vpPlus}
\centering
\end{subfigure}
\end{figure}

We start by expressing the causal set expectation value as a spacetime integral. The probability of sprinkling $n$ points in some region of spacetime, $\mathcal{R}$, is given by the Poisson distribution
\begin{equation}\label{eq:poisson_dist}
\mathbb{P}(n\text{ points in }\mathcal{R}) = \frac{(\rho\, \text{vol}(\mathcal{R}))^n}{n!} e^{-\rho\,  \text{vol}(\mathcal{R})} \;\;\; ,
\end{equation}
where $\rho$ is the density of the sprinkling, and $\text{vol}(\mathcal{R})$ is the spacetime volume of $\mathcal{R}$. For some small region, $\Delta \mathcal{R}$, the probability of sprinkling a single point is 
\begin{equation}
\mathbb{P}(1 \text{ point in }\Delta \mathcal{R}) = \rho\, \text{vol}(\Delta \mathcal{R}) e^{-\rho\,  \text{vol}(\Delta \mathcal{R})} \approx
\rho \, \Delta V \;\;\; ,
\end{equation}
where $\Delta V$ is the volume $\text{vol}(\Delta \mathcal{R})$. The probability of sprinkling a horizon $n$-molecule whose minimal element lies in a small region $\Delta \mathcal{R}_p$, about a point $p\in I^-(\mathcal{J})$, is 
\begin{align}\label{eq:prob_derivation}
\mathbb{P}(\text{horizon }n\text{-molecule beginning in }\Delta \mathcal{R}_p) & = \mathbb{P}(1 \text{ point in }\Delta \mathcal{R}_p) \nonumber
\\
& \times \mathbb{P}(n\text{ points in }I^+(p)\cap \mathcal{M}^-_+) \nonumber
\\
& \times \mathbb{P}(0\text{ points in }I^+(p)\cap \mathcal{M}^-_-) \nonumber
\\
& \approx \rho\,  \Delta V_p \, \frac{(\rho V_+(p))^n}{n!}e^{-\rho V_+(p)}e^{-\rho V_-(p)}  \nonumber
\\
& \approx \rho\,  \Delta V_p \, \frac{(\rho V_+(p))^n}{n!}e^{-\rho V(p)} \;\;\; ,
\end{align}
where $\Delta V_p$ is the small volume, $\text{vol}(\Delta \mathcal{R}_p)$, at $p$, and where we have defined the functions $V_{\pm}(p):=\text{vol}(I^+(p)\cap \mathcal{M}^-_{\pm})$ and $V(p):=V_+(p)+V_-(p)$. Figures~\ref{fig:vp} and~\ref{fig:vpPlus} illustrate these volumes.

The expected number of horizon $n$-molecules is a sum of the last line of~\eqref{eq:prob_derivation} over all $p\in I^-(\mathcal{J})$, in the limit that the small volumes $\Delta V_p$ go to zero. In this limit we replace the sum by an integral over all $p\in I^-(\mathcal{J})$ 
and obtain the following expression for the expected number of horizon $n$-molecules
\begin{equation}\label{eq:general_average}
\rho^{\frac{2-d}{d}} \langle \mathbf{H}_n \rangle = \rho^{\frac{2-d}{d}+1} \int_{I^-(\mathcal{J})} dV_p \frac{(\rho V_+(p))^n}{n!}e^{-\rho V(p)} \;\;\; ,
\end{equation}
where we have multiplied both sides by a factor of $\rho^{\frac{2-d}{d}}$.

\section{Rindler horizon in Minkowski space with a flat hypersurface} \label{sec:rindler}

The simplest case we can consider is an acceleration horizon in Minkowski space
with a flat spacelike hypersurface, which we can take to be a constant time surface in 
some inertial frame. The following heuristic argument supports the claim that the result in this simplest case will give us the 
leading term in the general case. Consider the $n=1$ case for 
definiteness. 
 The requirement that 
$p_-$ is maximal-but-one in $\mC^-$ means that it is close to $\Sigma$, and as $\rho \rightarrow \infty$ it gets closer.  The fact that $p_-$ lies in 
$I^-(\mJ)$ means that as $p_-$ approaches $\Sigma$, $p_-$ also gets closer to $\mH$, (see figure~\ref{fig:horizon_molecule}). 

We can see this tendency by inspecting the integrand of (\ref{eq:general_average}) in which the exponential will tend to suppress the integral in the region where $\rho V(p) \gg 1$. Indeed, the region where $\exp(-\rho V(p))$ is non-negligible is a small and decreasing subregion of $I^-(\mJ)$, immediately to the past of $\mJ$, converging on $\mJ$ as $\rho$ increases (see figure~\ref{fig:horizon_molecule}). 
 In the limit, the integral can therefore only depend on
geometric quantities at $\mJ$. On dimensional grounds, the only geometric quantity that can appear on the RHS of~\eqref{eq:general_area_result} is the 
area of $\mJ$ times a dimensionless constant, 
$a^{(d)}$, which is independent of the geometry. 	
Later, we will provide more evidence for this, but assuming it is true
we can determine the constant $a^{(d)}$ by considering the ``all-flat'' case of Minkowski space, flat $\Sigma$ and Rindler horizon $\mH$. We now turn to this
calculation. 

Consider $d$-dimensional Minkowski space with inertial coordinates 
$(x^0, x^1, y^\alpha )$, $\alpha = 2,3,\dots d-1$, and let $\Sigma$ be the hypersurface 
 $x^0= 0$. For this calculation we will employ the order reversed (past-future swapped) setup for convenience, so that the integral is over points $p\in I^+(\mathcal{J})$. $\mH$ is given by $x^0 = -x^1$. The region of integration is bounded by $\mH$ and by $x^0 = x^1$ and $x^0 > 0$. The integrand is independent of $y^\alpha$ and we have 
\begin{equation}\label{eq:flat_expectation_in_term_of_I}
\rho^{\frac{2-d}{d}} \langle \mathbf{H}_n \rangle = \int_{\mathcal{J}} d^{d-2}y \, I_n^{(d, flat)}(l) \, ,
\end{equation}
where the (dimensionless) function $I_n^{(d, flat)}(l)$  is
\begin{equation}\label{flatflatintegral}
I_n^{(d,flat)}(l) := \frac{l^{-(dn+2)}}{n!}
 \int_0^{\infty} dx^0 \int_{-x^0}^{x^0}dx^1 (\tilde{V}_+(x))^n e^{-\rho \tilde{V}(x)} \;\;\; ,
\end{equation}
where $l=\rho^{-1/d}$ is the discreteness length.
$\tilde{V}(x)$ is the $d$-dimensional volume of a solid null cone of height $x^0$,
\begin{equation}\label{eq:flat_cone_volume}
\tilde{V}(x) = \frac{S_{d-2}}{d(d-1)} \left( x^0 \right)^d \;\;\; ,
\end{equation}
where $S_d : = (d+1)\pi^{\frac{d+1}{2}}/\Gamma[\frac{d+1}{2}+1]$ is the volume of a unit $d$-sphere. Since the flat cone volume only depends upon $x^0$, we can  write it as a function of $x^0$ only.

The calculation of $\tilde{V}_+(x)$ is more complicated and we did not manage to determine a formula for general dimension $d$. For $d=2$, $\tilde{V}_+(x)$ is given by the following integral:
\begin{equation}
\tilde{V}_+(x) = \int_0^{\frac{1}{2}(x^0 - x^1)} d{x'}^0 \int_{x^1 - x^0 + {x^{'}}^0}^{-{x'}^0} d {x'}^1 \;\;\; .
\end{equation}
For $d\geq 3$, we can change to polar coordinates, $(R, \phi^I)$ (where $I=3,...,d-1$), in the $y^{\alpha}$ directions:
\begin{equation}\label{eq:y_polar_coordinates}
y^{\alpha} = R \, \zeta^{\alpha}(\phi) \;\;\; ,
\end{equation}
where $\zeta^{\alpha}(\phi)$ are the usual functions of the angular coordinates $\phi^I$. For $d=3$, it will be convenient below to sometimes use the coordinate $R$ instead of the coordinate $y^2$. In such cases, there will be a symmetry about the $x^A$ plane, so that one only needs to consider $R\geq 0$. $\tilde{V}_+(x)$ is given by the following integral:
\begin{align}\label{eq:all_flat_vplus_integral}
\tilde{V}_+(x) = & \int_0^{\frac{1}{2}(x^0 - x^1)} d{x'}^0 \, \int_{x^1 - x^0 + {x'}^0}^{- {x'}^0} d {x'}^1 \nonumber
\\
&  \int_0^{ \sqrt{ (x^0 - {x'}^0)^2 - ({x'}^1 - x^1)^2 } } d R \int_{\mathbb{S}^{d-3}} d \Omega_{d-3} \, R^{d-3} \;\;\; .
\end{align}
For $d=2,3$, and $4$, the above integrals give
\begin{align}
\label{eq:vplus_dtwo}
d=2 \, : \;\;\; \tilde{V}_+(x)= & \frac{1}{4}(x^0-x^1)^2  \;\;\; ,
\\
\label{eq:vplus_dthree}
d=3 \, : \;\;\; \tilde{V}_+(x)= & \frac{2}{3} {(x^0)}^3 \tan^{-1}\left( \sqrt{\frac{x^0-x^1}{x^0+x^1}} \right) \nonumber
\\
 -\frac{1}{9}(2 x^0 & -x^1) (x^0+ 2 x^1) \sqrt{(x^0-x^1)(x^0+x^1)}
\;\;\; ,
\\
\label{eq:vplus_dfour}
d=4 \, : \;\;\; \tilde{V}_+(x)= & \frac{\pi}{48}(x^0-x^1)^3 (5 x^0+3 x^1) \;\;\; .
\end{align}
For $d=2$ and $d=4$, the $l\rightarrow 0$ limit of $I_n^{(d, flat)}(l)$ can be be evaluated (using Watson's lemma~\cite{watson}) for any integer $n\geq 1$. One finds
\begin{align}
\lim_{l\rightarrow 0}I_n^{(2, flat)}(l)=a^{(2)}_n = & \frac{1}{2n+1} \;\;\; ,
\\
\lim_{l\rightarrow 0} I_n^{(4, flat)}(l)=a_n^{(4)} = & \frac{2^{8 n+1} }{3^{3 n+\frac{1}{2}}\sqrt{\pi } n!}\Gamma \left(n+\frac{1}{2}\right)\Bigg[ B_{\frac{3}{8}}(3 n+1,n+1) \nonumber
\\
& +B_{\frac{5}{8}}(n+1,3 n+1) -B_{\frac{1}{4}}(n+1,3 n+1)\Bigg] \;\; ,
\end{align}
where
\begin{equation}
B_z(a,b)\equiv\int_0^z ds \, s^{a-1} (1-s)^{b-1} \;\;\; .
\end{equation}
We can also evaluate the $l\rightarrow 0$ limit of $I_n^{(d, flat)}(l)$ for $n=1$ and $d=3$. One finds
\begin{align}
\lim_{l\rightarrow 0} I_1^{(2,flat)}(l)=a^{(2)}\equiv   a^{(2)}_1 &= \frac{1}{3} \;\;\; ,
\\
\lim_{l\rightarrow 0} I_1^{(3,flat)}(l)=a^{(3)} \equiv   a^{(3)}_1&  = \frac{1}{4}\left(\frac{3}{\pi} \right)^{\frac{2}{3}} \Gamma(\frac{5}{3}) \approx 0.218853 \;\;\; ,
\\
\lim_{l\rightarrow 0} I_1^{(4,flat)}(l)=a^{(4)} \equiv   a^{(4)}_1& = \frac{\sqrt{3}}{10} \approx 0.173205 \;\;\; ,
\end{align}
where we have included the $d=2$ and $d=4$ cases for completeness.

In all cases, the deviation from the limiting value tends to zero exponentially fast. For future reference, we comment here that were the integral over $x^0$ in 
(\ref{flatflatintegral}) to be cut off at any finite upper limit, $\tau$ say, this would not affect the value of the $l \rightarrow 0 $ limit because the difference will vanish exponentially fast in the limit. This will be important in the general curvature case below. 

\section{General curvature}\label{sec:horizon_area}

We  turn to the general case and provide a more detailed argument for 
why the flat result above gives the limiting value of the mean number of molecules per unit horizon area. 
  Since we can take the discreteness length, $l$, to be as small as we like in \eqref{eq:general_average}, 
we can take it to be much smaller than the curvature scales of the spacetime and of the two surfaces $\mathcal{H}$ and $\Sigma$. 
Concretely, we assume there is a length $\tau$ such that $l<<\tau<< L_G$
where $L_G$ denotes the smallest geometric length scale in our setup.  
The ratio $\varepsilon:=\tau / L_G << 1$ will be useful as an expansion parameter. 
Note that for Causal Set Theory, this is the physically relevant regime because the continuum approximation is only valid when the curvature length scales involved in the problem are much larger than the discreteness scale, $l$. 

\subsection{Local geometric invariants and Florides-Synge Normal Coordinates}\label{sec:Florides-Synge Normal Coordinates}

$\Sigma$ can be considered to be a member of a family of hypersurfaces given by 
$S_{\Sigma}(z) =$ $constant$ where $S_{\Sigma}(z)$ is a spacetime function that is 
zero on $\Sigma$, and increases to the past. Here, $z^a$ ($a=0,1,...,d-1$) are coordinates on $\mathcal{M}$. The components of the normal covector are given by
\begin{equation}\label{eq:normal_covector_definition}
n_{a} := (-g^{bc}\partial_b S_{\Sigma}\partial_c S_{\Sigma})^{-1/2} \partial_a S_{\Sigma} \, ,
\end{equation}
The components of the normal vector are  $n^a = g^{ab}n_b$ and it is 
future pointing. The projector $h^a_{\; b}:=\delta^a_b + n^an_b$, on $\Sigma$, projects vectors onto the tangent space of $\Sigma$. The extrinsic curvature tensor for $\Sigma$ is
\begin{equation}\label{eq:extrinsic_curvature_definition}
K_{ab}:= n_{ c \, ; \, d }h^c_{\; a} h^d_{\; b} \;\;\; .
\end{equation}
The trace of the extrinsic curvature is
\begin{equation}\label{eq:extrinsic_curvature_scalar_definition}
K := g^{ab}K_{ab} = h^{ab}K_{ab} \;\;\; ,
\end{equation}
where  the index of $h^a_{\; b}$ is raised using the inverse metric $g^{ab}$.

\begin{figure}[t]
\caption{\footnotesize{An illustration of the $x^A = (x^0,x^1)$ plane  though a point $q$ in $\mJ$, with the vectors used in our setup. $n$ is normal to $\Sigma$, $k$ is normal to $\mathcal{H}$, $l$ is orthogonal to the coordinate vectors $\partial /\partial y^{\alpha}$ and satisfies $k.l=-1$, and $m$ is tangent to $\Sigma$. All these vectors are orthogonal to $\mJ$.}}
\includegraphics[trim={0 4cm 0 4cm},clip,width=12cm]{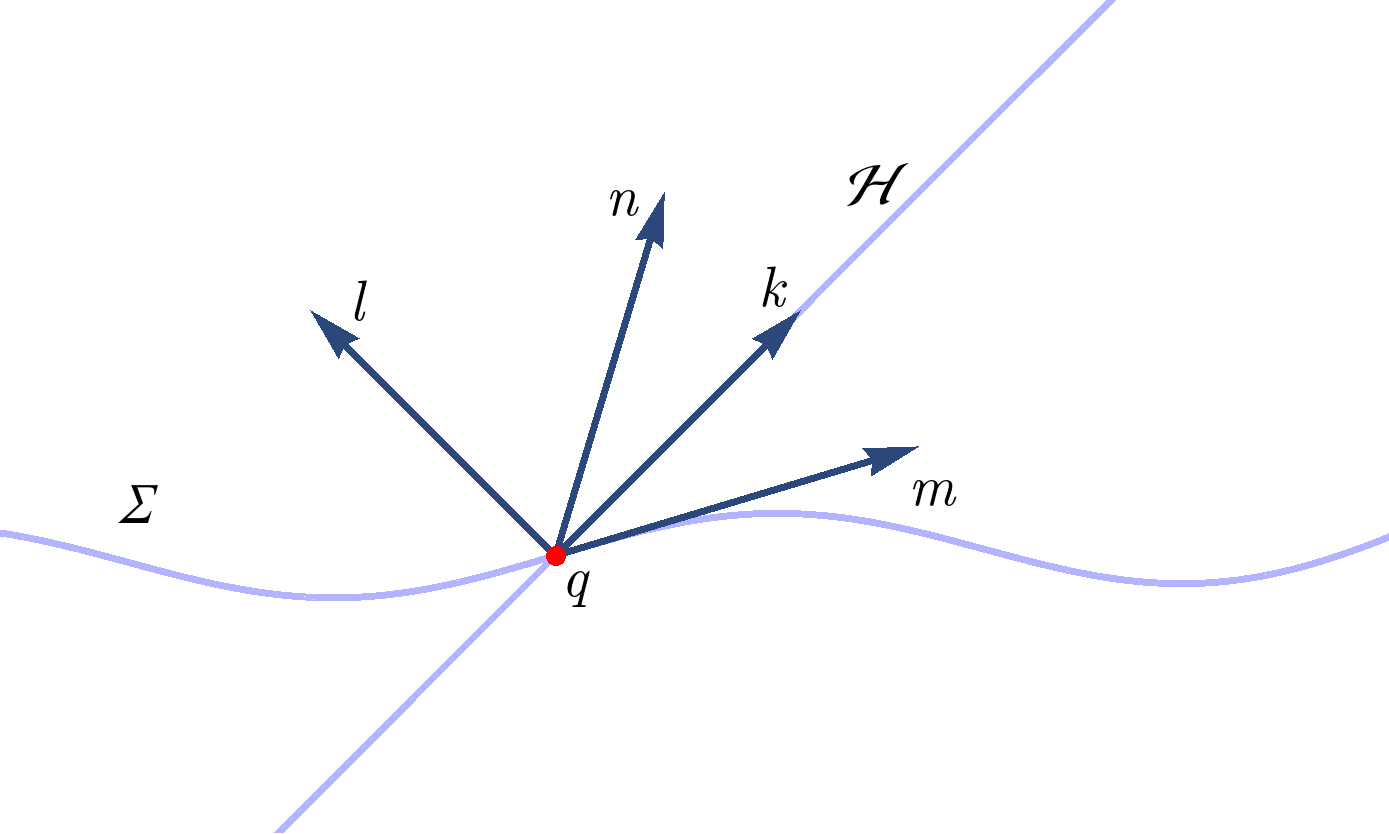}\label{fig:vectors}
\centering
\end{figure}

Similarly  $\mathcal{H}$ can be considered to be a member of a family of hypersurfaces. We fix the normalisation of the future-directed normal vector $k$ to  $\mathcal{H}$ by $n\cdot k=-1/\sqrt{2}$. $k$
 is tangent to the null geodesic generators of 
$\mathcal{H}$.  We assume that exactly one such null geodesic generator passes through any given point on $\mathcal{J}$,\footnote{This assumption will not necessarily hold at all points on $\mJ$ because of the existence of caustics. However, the points on $\mJ$ at which it fails are a set of measure zero and so will not 
affect our results, which end up being integrals over $\mJ$.}   and so we can use any coordinates $y^{\alpha}$, $\alpha = 2,3,\dots d-1$, on $\mathcal{J}$, to label the generators.  We can  uniquely define a second future-directed null vector $l$, within some neighbourhood about $\mathcal{J}$ within $\mathcal{H}$, as that which satisfies $l.k=-1$, and is orthogonal to every coordinate vector $\partial/\partial y^{\alpha}$. We  define the tensor $\sigma^a_{\; b}: = \delta^a_b + k^al_b + l^ak_b$, on $\mathcal{H}$, which projects onto the tangent space of $\mathcal{J}$. Note also that $\sigma^a_{\; b} = h^a_{\; b} -m^a m_b$, where $m$ is a spacelike vector, normalised as $m.m=1$, defined (only on $\mathcal{J}$) as $m:=\sqrt{2}\left[k+(k.n)n\right]$. See figure~\ref{fig:vectors} for an illustration of these vectors. The null expansion scalar is  defined as
\begin{equation}\label{eq:theta_definition}
\theta := k_{a ; b} \sigma^{ab} \;\;\; ,
\end{equation}
where the index of $\sigma^a_{\; b}$ has been raised by $g^{ab}$.

We will need coordinates tailored to our geometrical setup,  focussed on the  intersection $\mJ$, its neighbourhood and the normal vectors, $n$ and $k$ to $\Sigma$ and $\mH$, respectively. Florides-Synge Normal Coordinates (FSNC's) can be constructed in a tubular neighbourhood about a submanifold of any co-dimension in any Riemannian, or pseudo-Riemannian, manifold \cite{Florides2}. Here we consider the specific case of FSNC's based around the co-dimension 2 spacelike submanifold $\mathcal{J}$ and tailored to $\Sigma$ and $\mH$. 

For $d>2$, one can construct FSNC's $z^a = (x^A, y^{\alpha})$ ($a=0,...,d-1$, $A=0,1$, and $\alpha=2,...,d-1$), within a small enough tubular neighbourhood $\mathcal{N}$, as follows. First, we choose any coordinates $y^{\alpha}$ on $\mathcal{J}$ (in general one will have an atlas of charts on $\mathcal{J}$). Next, pick any smooth orthogonal frame of vectors for each point $q\in\mathcal{J}$, such that two of the vectors in each frame are orthogonal to $\mathcal{J}$ (the transverse directions). We choose these transverse vectors to be $n$ and $m$ as defined above. Note that $n.m=0$, so that $m$ lies within the part of the tangent space of $\Sigma$ that is orthogonal to the tangent space of $\mathcal{J}$.

Consider, from each point $q\in\mathcal{J}$ with coordinates $y^{\alpha}$, sending out a two parameter family of geodesics with tangent vectors  $v=x^0 n + x^1 m$ on $\mJ$. The point $p$ which is affine parameter distance $1$ away from $q$ in $\mJ$ along the geodesic with tangent vector  $x^0 n + x^1 m$ has FSNC's  
$z^a := (x^A, y^{\alpha})$. For $x^0$ and $x^1$ small enough, this is well defined. 

In FSNC's the submanifold $\mathcal{J}$ is described by the equation $x^A=0$ and the horizon $\mathcal{H}$ is given by the equation
\begin{equation}
x^0 = x^1 \;\;\; ,
\end{equation}
within the tubular neighbourhood $\mathcal{N}$.
The generator of $\mH$ through $q$ in $\mJ$ with coordinates $y^\alpha$ is
described by the curve $z^a(\lambda)=(\lambda/\sqrt{2},\lambda/\sqrt{2},y^{\alpha})$, where $\lambda$ is the affine parameter on the geodesic. 
When $d=2$  there are no $y^{\alpha}$ coordinates, and the coordinates $x^A$ are Riemann Normal Coordinates (RNC's)  about the intersection $\mathcal{J}$, which is a point in $d=2$. In what follows we will mostly assume that $d>2$, and we will only restrict to the simpler case of $d=2$ when necessary.

We have the coordinate conditions 
\begin{equation}
g_{a B}(x,y)x^B = \eta_{aB}x^B \;\;\; .
\end{equation}
The metric $g_{ab}(x,y)$ can be expanded about $\mathcal{J}$, i.e. in small $x^A$, as
\begin{align}
g_{AB}(x,y) & = \eta_{AB} +O(x^2) \;\;\; ,
\\
g_{A\beta}(x,y) & = g_{A\beta,C}(0,y)x^C +O(x^2) \;\;\; ,
\\
g_{\alpha\beta}(x,y) & = \sigma_{\alpha\beta}(y)+ g_{\alpha\beta,C}(0,y)x^C +O(x^2) \;\;\; ,
\end{align}
where $\sigma_{\alpha\beta}(y)$ is the induced metric on $\mathcal{J}$. The metric determinant can be expanded as
\begin{equation}\label{eq:metric_determinant_expansion}
\sqrt{-g(x,y)} = \sqrt{\sigma(y)}+ O(x) \;\;\; ,
\end{equation}
where $\sigma(y)$ is the determinant of the metric $\sigma_{\alpha\beta}(y)$.

\subsection{Reducing to a local integral}

Consider the tubular neighbourhood, $\mathcal{N}\supset\mathcal{J}$, in which the FSNC's have been constructed, and define the region
\begin{equation}
\mathcal{R}_{\tau}: = \lbrace \, p\in I^-(\mathcal{J})\cap \mathcal{N} \, : \, -\tau < x^0(p) < 0  \rbrace \;\;\; ,
\end{equation}
where $x^A(p)$ are the transverse coordinates of the point $p$. 
$\tau$ is the middle scale of the hierarchy of scales,  $l<<\tau<< L_G$,   discussed above, and is assumed to be small enough that this region is inside $\mathcal{N}$, where the FSNC's are defined.

Additionally, define the complement $\overline{\mathcal{R}}_{\tau} := I^-(\mathcal{J})\setminus \mathcal{R}_{\tau}$. The integral in~\eqref{eq:general_average} can be split into a part over $\mathcal{R}_{\tau}$ and a part over $\overline{\mathcal{R}}_{\tau}$. 

We need the integral over $\overline{\mathcal{R}}_{\tau}$ to tend to zero faster than any power of $l$  so that we can ignore its contribution to the result in what follows. 
This will be so if the region $\overline{\mathcal{R}}_{\tau}$  has finite volume,  for example if $\mM$ itself has finite volume to the past of $\Sigma$,  since 
\begin{align}
\left| \int_{\overline{\mathcal{R}}_{\tau}} dV_p  V_+(p)^n e^{-\rho V(p)} \right|
& \leq \left| \int_{\overline{\mathcal{R}}_{\tau}} dV_p  V(p)^n e^{-\rho V(p)} \right| \nonumber 
\\
& \leq \text{max}_{\overline{\mathcal{R}}_{\tau}}\big[V(p)^n e^{-\rho V(p)}\big]\int_{\overline{\mathcal{R}}_{\tau}} dV_p  \nonumber
\\
& = V_{\text{min}}^n e^{-\rho V_{\text{min}}} \text{vol}(\overline{\mathcal{R}}_{\tau}) \;\;\; .
\end{align}
In the last line we have defined $V_{\text{min}}$ as the minimum value of $V(p)$ for $p\in\overline{\mathcal{R}}_{\tau}$. This minimum value will be achieved at some $p$ on the future spacelike boundary of $\overline{\mathcal{R}}_{\tau}$ and the integral over $\overline{\mathcal{R}}_{\tau}$ is exponentially suppressed. 

If $\mathcal{M}$ does not have finite volume to the past of $\Sigma$, the integral over $\overline{\mathcal{R}}_{\tau}$ can still be exponentially suppressed as $l\rightarrow 0$. For example, this is the case for any $\tau >0$ in Minkowski space. We give a plausibility argument why it will be true more generally. We assume that the level sets of $V(p)$, for all $p\in I^-(\mathcal{J})$, foliate the sub-spacetime $I^-(\mathcal{J})$ into compact, measurable leaves. That is, for any $v>0$, the set of all $p\in I^-(\mathcal{J})$ such that $V(p)=v$ is some compact measurable set, $\Sigma_v$. Given this assumption, we can use $v$ as a ``time coordinate'' on $I^-(\mathcal{J})$ and bound
\begin{align}
\left| \int_{\overline{\mathcal{R}}_{\tau}} dV_p  V_+(p)^n e^{-\rho V(p)} \right|
& \leq \left| \int_{\overline{\mathcal{R}}_{\tau}} dV_p  V(p)^n e^{-\rho V(p)} \right| \nonumber 
\\
& = \left|  \int_{v_0}^{\infty}dv \, v^n e^{-\rho v}  \, f(v) \right| \nonumber \;\;\; ,
\end{align}
where $v_0$ is the minimum value $V(p)$ takes for all $p\in \overline{\mathcal{R}}_{\tau}$, and where $f(v)$ is the integral of the volume measure $\sqrt{-g}$ over $\Sigma_v$. Following the proof of Watson's lemma~\cite{watson}, we can bound the integral on the last line if we assume that $|f(v)|$ has at most exponential growth as $v\rightarrow\infty$, i.e. $|f(v)|\leq C e^{C' v}$, for all $v \geq v_0$ (where $C$ and $C'$ are constants that are independent from $\rho$). We have that 
\begin{align}
\left| \int_{v_0}^{\infty}dv \, v^n f(v) e^{-\rho v} \right|
& \leq  \int_{v_0}^{\infty}dv \, v^n |f(v)| e^{-\rho v}  \nonumber 
\\
& \leq  C \int_{v_0}^{\infty}dv \, v^n e^{(C'-\rho) v} \nonumber
\\
& =  e^{(C'-\rho)v_0} \left[ \frac{C {v_0}^n}{C'-\rho} + O\left( (C'-\rho)^{-2}\right)\right] \;\;\; .
\end{align}
We leave it as an open problem to determine the class of spacetimes for which $|f(v)|$ has at most exponential growth. 

We will henceforth assume that the integral over $\overline{\mathcal{R}}_{\tau}$ is exponentially suppressed in the limit and write the  expected value as
\begin{equation}\label{eq:general_avg_R}
\rho^{\frac{2-d}{d}} \langle \mathbf{H}_n \rangle = \rho^{\frac{2-d}{d}+1} \int_{\mathcal{R}_{\tau}} dV_p \frac{(\rho V_+(p))^n}{n!}e^{-\rho V(p)} + \cdots \;\;\; ,
\end{equation}
where ``$\cdots$" denotes terms that tend to zero exponentially fast in the limit. 
The region $\mathcal{R}_{\tau}$ lies within the region of validity of our FSNC's, and hence we can  write the expectation value explicitly in terms of our FSNC's:
\begin{align}\label{eq:general_avg_fsnc}
\rho^{\frac{2-d}{d}} \langle \mathbf{H}_n \rangle = &  \frac{\rho^{\frac{2-d}{d}+1+n}}{n!}
\int_{\mathcal{J}} d^{d-2}y \int_{-\tau}^0 dx^0 \int_{x^0}^{-x^0}dx^1 \sqrt{-g(x,y)}(V_+(x,y))^n e^{-\rho V(x,y)}  \nonumber
\\
& + \cdots \;\;\; ,
\end{align}
where we have written the volume $V(p)$ as a function of the coordinates $x^A$ and $y^{\alpha}$.

Then, 
\begin{equation}\label{eq:expectation_in_term_of_I}
\rho^{\frac{2-d}{d}} \langle \mathbf{H}_n \rangle = \int_{\mathcal{J}} d^{d-2}y \sqrt{\sigma(y)}\, I_n^{(d)}(y; l, \tau) + \cdots \;\;\; ,
\end{equation}
where we have defined 
\begin{equation}
I_n^{(d)}(y; l, \tau) := \frac{l^{-(dn+2)}}{n!}
 \int_{-\tau}^0 dx^0 \int_{x^0}^{-x^0}dx^1 \sqrt{-\frac{g(x,y)}{\sigma(y)}}(V_+(x,y))^n e^{-\rho V(x,y)} \;\;\; .
\end{equation}
The factor $\frac{g(x,y)}{\sigma(y)}$ makes $I_n^{(d)}(y; l, \tau)$ a scalar on $\mathcal{J}$ 
and we rewrite it in a coordinate free notation as $I_n^{(d)}(q;l,\tau)$, where $q\in\mathcal{J}$.
 
$I_n^{(d)}(q;l,\tau)$ is uniquely specified given the spacetime, $\Sigma$, $\mH$, the point $q \in\mathcal{J}$ and 
the lengths $l$ and $\tau$. As $l$ tends to zero, the region in which the integrand in $I_n^{(d)}(q;l,\tau)$ is non-negligible converges on 
the point $q$, and we conclude that $I_n^{(d)}(q;l,\tau)$ has a small $l$ expansion of the form
\begin{equation}\label{eq:general_expansion_of_I_function}
I_n^{(d)}(q;l,\tau) = a_n^{(d)} + l \sum_i b_{n,i}^{(d)} \mathcal{G}_i(q) +O(l^2) \;\;\; .
\end{equation}
Here $a_n^{(d)}$ and $b_{n,i}^{(d)}$ are constants that only depend upon the dimension $d$, and the integer $n$. The set $\lbrace \mathcal{G}_i(q) \rbrace$ is the largest set of mutually independent geometric scalars of length dimension $L^{-1}$ evaluated at $q$, and the subscript $i$ simply indexes this set. For example, $\mathcal{G}_1(q)$ could be the extrinsic curvature scalar $K$ evaluated at $q$, and $\mathcal{G}_2(q)$ could be the null expansion $\theta$ at $q$. In the above equation, the sum over $i$ runs over the whole set $\lbrace \mathcal{G}_i(q) \rbrace$. Note that the set $\lbrace \mathcal{G}_i(q) \rbrace$ is not unique. Relations between scalars, such as the contracted form of the Gauss-Codazzi equations~\cite{Poisson:2009pwt}, mean that we may have a choice as to which scalars to include in the set $\lbrace \mathcal{G}_i(q) \rbrace$. Its cardinality, however, is unique. Given~\eqref{eq:general_expansion_of_I_function} we have that
\begin{equation}
\lim_{l\rightarrow 0} I_n^{(d)}(q;l,\tau) = a_n^{(d)} \;\;\; ,
\end{equation}
which implies our claim ~\eqref{eq:general_area_result_n}. 

\subsection{Expansion of $I_n^{(d)}(q; l, \tau)$}

We will examine the small $l$ expansion in more detail. It will be convenient to again switch to an order-reversed setup. In this case, $\mathcal{H}$ is given by $x^0 = - x^1$, and the points $p$ of each horizon molecule lie in the region $I^+(\mathcal{J})$. We also take $V_+(x,y)$ and $V(x,y)$ to represent the volumes of the corresponding order reversed regions. We will also order reverse the normal vectors $k$ and $n$, so that they are  past-pointing\footnote{By order reversing the vectors we ensure that the constants $a^{(d)}_n$ and $b^{(d)}_{i,n}$ (to be determined in the following sections) have the same form (in terms of $d$ and $n$) as those in our original geometric setup with future-pointing vectors.}. For a visualisation of this order-reversed setup, imagine reversing the time-axis of figures~\ref{fig:vectors},~\ref{fig:vp}, and~\ref{fig:vpPlus}. The function $I_n^{(d)}(q; l, \tau)$ is  given by
\begin{equation}
I_n^{(d)}(q; l, \tau) = \frac{l^{-(dn+2)}}{n!}
 \int_0^{\tau} dx^0 \int_{-x^0}^{x^0}dx^1 \sqrt{-\frac{g(x,y)}{\sigma(y)}}(V_+(x,y))^n e^{-\rho V(x,y)} \;\;\; .
\end{equation}
We are also free too choose the coordinates $y^{\alpha}$ on $\mathcal{J}$, and hence we can choose RNC's (within $\mathcal{J}$) centred about $q\in\mathcal{J}$. The expressions $g(x,y)$, $\sigma(y)$, $V_+(x,y)$, and $V(x,y)$, that depend on the coordinates $y^{\alpha}$, are  all evaluated at $y^{\alpha}=0$, and hence we will drop that argument entirely. We also have that $\sigma(0)=1$ in these RNC's on $\mathcal{J}$ and 
\begin{equation}
I_n^{(d)}(q; l, \tau) = \frac{l^{-(dn+2)}}{n!}
 \int_0^{\tau} dx^0 \int_{-x^0}^{x^0}dx^1 \sqrt{-g(x)}(V_+(x))^n e^{-\rho V(x)} \;\;\; .
\end{equation}
We can introduce spacetime RNC's $Z^a = (X^A , Y^{\alpha})$, in a neighbourhood $\mathcal{U}$ about $q$, such that $X^A = x^A$, and such that the coordinate vectors $\partial/\partial Y^{\alpha} = \partial/ \partial y^{\alpha}$ at $q$. This ensures that the determinant of the metric, evaluated at $q$ ($y^{\alpha} = Y^{\alpha} = 0$), has the same form in terms of the coordinates $x^A$ and $X^A$. We can  write
\begin{equation}\label{eq:I_function_RNCs}
I_n^{(d)}(q; l, \tau) = \frac{l^{-(dn+2)}}{n!}
 \int_0^{\tau} dX^0 \int_{-X^0}^{X^0}dX^1 \sqrt{-g(X)}(V_+(X))^n e^{-\rho V(X)} \;\;\; ,
\end{equation}
in terms of the RNC's, $Z^a$, about $q$.

The determinant $g(X)$ can be expanded in small $X^A$ relative to the curvature scales of the spacetime at $q$:
\begin{equation}
\sqrt{-g(X)} = 1 -\frac{1}{6} R_{AB} X^A X^B + O(Z^3) \;\;\; ,
\end{equation} 
where the Ricci tensor $R_{ab}$ has been evaluated at $q$, and we only have a contraction over the indices $A,B=0,1$ as $Y^{\alpha}=0$. $R_{ab}$ has length dimensions of $L^{-2}$, and we can define a dimensionless tensor $\hat{R}_{ab} := {L_G}^2 R_{ab}$, using $L_G$ (the smallest geometric length scale from our setup). We can also rewrite the above expression in terms of dimensionless coordinates $\hat{Z}^a : = Z^a / \tau$:
\begin{align}
\sqrt{-g(\tau \hat{X})} & = 1 -\frac{1}{6}\left(\frac{\tau}{L_G}\right)^2 \hat{R}_{AB} \hat{X}^A \hat{X}^B + O(Z^3) 
\\
& = 1 -\frac{1}{6}\varepsilon^2 \hat{R}_{AB} \hat{X}^A \hat{X}^B + O(\varepsilon^3) \;\;\; .
\end{align}
In this way we can see that the correction $\frac{1}{6} R_{AB} X^A X^B$ is $O(\varepsilon^2)$. Recall that $\varepsilon=\tau/L_G << 1$. We have also written the higher order correction as $O(\varepsilon^3)$. From this point onwards, it will be more convenient to express higher order corrections in terms of $\varepsilon$.

We  turn our attention to the volumes $V(X)$ and $V_+(X)$. Given a point $p$ that has coordinates ${Z_p}^a=({X_p}^0,{X_p}^1,0)$, these volumes can be written as
\begin{align}
V(X_p) & = \int_{\mathcal{R}_p} d^dZ \sqrt{-g(Z)} \;\;\; ,
\\
V_+(X_p) & = \int_{\mathcal{R}_{p,+}} d^dZ \sqrt{-g(Z)} \;\;\; .
\end{align}
where $\mathcal{R}_p:=I^-(p)\cap I^+(\Sigma)$, and $\mathcal{R}_{p,+} := \mathcal{R}_p\cap I^-(\mathcal{H})$. The metric determinant can be expanded in small $\varepsilon$ as
\begin{equation}\label{eq:general_metric_det_expansion}
\sqrt{-g(Z)} = 1 -\frac{1}{6} R_{ab} Z^b Z^b + O(\varepsilon^3) \;\;\; .
\end{equation}

The past boundaries of the regions $\mathcal{R}_p$ and $\mathcal{R}_{p,+}$ are subregions of the surface $\Sigma$. The future boundary of $\mathcal{R}_p$ is some subregion of the past lightcone of $p$, denoted here by $\mathcal{P}:= \partial I^-(p)$, and the future boundary of $\mathcal{R}_{p,+}$ is made up of a subregion of $\mathcal{P}$ and a subregion of $\mathcal{H}$.

In any specific spacetime setup below, we will only consider surfaces $\Sigma$, $\mathcal{P}$, and $\mathcal{H}$, that can be described in the neighbourhood $\mathcal{U}$ by twice differentiable functions ${X_{\Sigma}}^0 (Z^i)$, ${X_{\mathcal{P}}}^0 (Z^i)$, and ${X_{\mathcal{H}}}^0 (Z^i)$ ($i=1,...,d-1$)~\footnote{We actually do not require the function ${X_{\mathcal{P}}}^0 (Z^i)$ to be twice differentiable at $p$.}. That is, functions from the spatial coordinates $Z^i$ to the time coordinate $X^0$. In the spacetime setups considered below we will calculate $V(X_p)$ and $V_+(X_p)$, and our calculations suggest expansions of the form
\begin{align}\label{eq:expansion_of_volumes_in_paramters}
V(X_p ) & = \tilde{V}(X_p) \left( 1 + \sum_i \mathcal{G}_i(q) f_i(X_p) + O(\varepsilon^2) \right) \;\;\; , \nonumber
\\
V_+(X_p ) & = \tilde{V}_+(X_p) \left( 1 + \sum_i \mathcal{G}_i(q) f_{+,i}(X_p) + O(\varepsilon^2) \right) \;\;\; ,
\end{align}
where $\tilde{V}(X_p)$ and $\tilde{V}_+(X_p)$ are the volumes from the {all-flat} case, considered in section \ref{sec:rindler}.

The functions $f_i(X_p)$ and $f_{+,i}(X_p)$ must have length dimensions $L$. Equivalently, we say that the functions must be homogeneous of degree 1, i.e. $f_i( \lambda X_p) = f_i(X_p)$ and $f_{+,i}( \lambda X_p) = f_{+,i}(X_p)$. We have also written the next order correction in terms of $\varepsilon$. This correction will likely involve scalars of length dimension $L^{-2}$, and homogeneous functions of the coordinates ${X_p}^A$ of degree 2.  It seems plausible that one could rigorously prove the expansions in \eqref{eq:expansion_of_volumes_in_paramters} given the assumption that the surfaces $\Sigma$, $\mathcal{P}$, and $\mathcal{H}$ are twice differentiable, and that the metric can be expanded as in~\eqref{eq:general_metric_det_expansion}. 

The above volume expansions actually imply that $I_n^{(d)}(q; l, \tau)$ has a small $l$ expansion of the form~\eqref{eq:general_expansion_of_I_function}. To show this we must use the volume expansions to expand $I_n^{(d)}(q; l, \tau)$ in $\varepsilon$. We begin by expanding the different parts of the integrand in~\eqref{eq:I_function_RNCs} in $\varepsilon$:
\begin{align}
\sqrt{-g(X)} & = 1 + O(\varepsilon^2) \;\;\; , 
\\
(V_+(X))^n & = \tilde{V}_+(X)^n \left( 1 + n \sum_i \mathcal{G}_i(q) f_{+,i}(X) + O(\varepsilon^2) \right) \;\;\; , 
\\
e^{-\rho V(X)} & = e^{- \rho \tilde{V}(X)} \left( 1 - \rho \tilde{V}(X)  \sum_i \mathcal{G}_i(q) f_i(X) + \rho \tilde{V}(X)O(\varepsilon^2) \right) \;\;\; ,
\end{align}
where we have removed the subscript $p$ from the coordinates $X^A$. Since the flat cone volume only depends upon $X^0$ we can take it out of the integral over $X^1$ in $I_n^{(d)}(q; l, \tau)$. We  have
\begin{align}\label{eq:I_function_first_order_expansion_with_general_scalars}
I_n^{(d)}(q; l, \tau) & = \frac{l^{-(dn+2)}}{n!}
 \int_0^{\tau} dX^0 e^{-\rho \tilde{V}(X^0)} \Bigg\lbrace \int_{-X^0}^{X^0}dX^1 \tilde{V}_+(X)^n \nonumber
 \\
& +  \sum_i \mathcal{G}_i(q) \Bigg[ n \int_{-X^0}^{X^0} dX^1 \tilde{V}_+(X)^n f_{+,i}(X) \nonumber
\\
& - \rho\tilde{V}(X^0)  \int_{-X^0}^{X^0} dX^1 \tilde{V}_+(X)^n f_i(X) \Bigg] \nonumber
\\
& + \int_{-X^0}^{X^0} dX^1 \tilde{V}_+(X)^n \Big( O(\varepsilon^2) + \rho \tilde{V}(X^0)O(\varepsilon^2)  \Big)  \Bigg\rbrace \;\;\; .
\end{align}
The integral in the first line equals the integral $I_n^{(d, flat)}(l)$ from section \ref{sec:rindler} up to a difference which vanishes exponentially fast in the limit 
as per the comment at the end of that section.

The $X^1$ integrals in lines 2 and 3 of~\eqref{eq:I_function_first_order_expansion_with_general_scalars} both have length dimensions $L^{dn+2}$, and they both only depend upon $X^0$. Therefore, they must evaluate to functions of the form
\begin{equation}
C (X^0)^{dn+2} \;\;\; ,
\end{equation}
for some constant $C$. This fact, together with Watson's lemma~\cite{watson}, mean that the expression in square brackets in lines 2 and 3 of~\eqref{eq:I_function_first_order_expansion_with_general_scalars} evaluates to a term of the form $C' l$, for some constant $C'$, as $l\rightarrow 0$. Similarly, the $O(\varepsilon^2)$ corrections in line 4 of~\eqref{eq:I_function_first_order_expansion_with_general_scalars} tend to a function of order $O(l^2)$ as $l\rightarrow 0$. We therefore have the small $l$ expansion
\begin{equation}
I_n^{(d)}(q;l,\tau) = a_n^{(d)} + l \sum_i b_{n,i}^{(d)} \mathcal{G}_i(q) +O(l^2) \;\;\; ,
\end{equation}
where $a_n^{(d)}$ have been shown to be the numbers given in section \ref{sec:rindler}. 
The explicit form of the constants $b^{(d)}_{n,i}$ can  be determined using geometric setups with non-zero scalars $\mathcal{G}_i(q)$. From~\cite{Khetrapal:2012ux} we do not expect the curvature of $\mathcal{P}$ to contribute at first order, and our explicit calculations in the next section are consistent with this.

\section{First order corrections}\label{sec:first_order_corrections}

In this section we will explore the $O(l)$ term in the small $l$ expansion of $I_n^{(d)}(q; l, \tau)$. After determining its exact form, we will be able to construct causal set expressions for extracting more geometrical information about the surfaces $\mathcal{H}$ and $\Sigma$.
In the next subsection we will explicitly write down the set of independent scalars $\lbrace\mathcal{G}_i(q) \rbrace$, and in the following section we will use specific setups, \textit{a la} Gibbons and Solodukhin~\cite{Gibbons:2007nm}, to determine the constants $b^{(d)}_{n,i}$.

\subsection{General form of the expansion}

To find all the independent scalars of dimension $L^{-1}$, we consider all possible first derivatives of vectors and tensors that depend upon the basic dimensionless geometrical objects at $\mJ$. We have the metric $g$, the normal vector $n$ to $\Sigma$, the normal vector $k$ to $\mH$, the spacelike vector $m$, and the null vector $l$ constructed from these, all as described in section \ref{sec:Florides-Synge Normal Coordinates}. A systematic process of taking first derivatives of these and forming scalars by contracting gives three independent scalars on $\mJ$ of length dimension $L^{-1}$:  $\theta$ the null expansion of $\mH$, $K$ the trace of the extrinsic curvature of $\Sigma$,  and the component 
$K_{11} : = K_{ab} m^a m^b$ of the extrinsic curvature. See appendix~\ref{app:determining_the_set_of_scalars} for more details. We therefore expect $I_n^{(d)}(q; l, \tau)$ to have the small $l$ expansion
\begin{equation}\label{eq:integral_expansion_claim}
I_n^{(d)}(q; l, \tau) =  a^{(d)}_n + \left ( b^{(d)}_{1,n}K+b^{(d)}_{2,n}K_{11} + b^{(d)}_{3,n}\theta \right) l +O(l^2)  \;\;\; ,
\end{equation}
where $\theta$, $K_{11}$, and $K$ are evaluated at $q\in\mathcal{J}$. Assuming this form we can determine the constants by calculating the expansion of $I_n^{(d)}(q; l, \tau)$ for specific setups. %

For $d=2$ there is no null expansion $\theta$. Additionally, $K=K_{11}$, and hence these two scalars are not independent. In that case, we expect a small $l$ expansion of the form
\begin{equation}\label{eq:integral_expansion_claim_2d}
I_n^{(2)}(q; l, \tau) =  a^{(2)}_n +  b^{(2)}_{n}K \, l +O(l^2)  \;\;\; .
\end{equation}

We will set up the evaluation of the constants $b^{(d)}_{i,n}$, for general dimension $d$ but we will only find the final expressions for $d=2$ and $d=4$ and leave the determination of closed form expressions for $b^{(d)}_{i,n}$ 
for future work.

It is worth commenting on the appearance of $\theta$ in the above expansion for $d\geq 3$. One may worry that this is not geometric, as $\theta$ depends on the choice of parameter $\lambda$ along the null geodesics ruling $\mathcal{H}$. Here we have chosen a particular parameter by requiring that the parameter is affine, and that $n.k=-1/\sqrt{2}$. If we were to scale our affine parameter, the value of $\theta$ would scale in the same way. In the calculations below one can see that the coefficient $b^{(d)}_{3,n}$ would scale in the inverse way, such that the combination $b^{(d)}_{3,n}\theta$ remains unchanged. We should therefore think of the combination $b^{(d)}_{3,n}\theta$ as the truly geometric quantity. 

\subsection{Determining the constants}

\subsubsection{$b^{(d)}_{1,n}$}

To determine the constant $b^{(d)}_{1,n}$ we choose a setup such that $K_{11}=\theta=0$. Specifically, we take the spacetime $\mathbb{M}^d$ ($d>2$), with coordinates $Z^a=(X^A,Y^{\alpha})$, as in our all-flat calculations. It will be more convenient to leave the determination of $b^{(d)}_{1,n}$, for $d=2$, to the next section. For convenience, we will also stick to a order reversed geometric setup during the calculations of the three constants. We wish to find the first order correction to the function $I_n^{(d)}(q; l, \tau)$, evaluated at some point $q$, which we take to be the origin, $Z^a=0$. Given the order reversed setup, the null surface $\mathcal{H}$ is given by
\begin{equation}
X^0=-X^1\;\;\; ,
\end{equation}
which ensures that $\theta=0$.

In this setup the spacelike surface $\Sigma$ is given by the zeroes of the function
\begin{equation}
S_{\Sigma}(Z) = X^0 - a\,  R^2 \;\;\; ,
\end{equation}
where $R=\sqrt{\delta_{\alpha\beta}Y^{\alpha}Y^{\beta}}$ is the radius in the $Y^{\alpha}$ directions that was introduced above. The free parameter $a>0$ controls how curved $\Sigma$ is. Note that the above equation only describes $\Sigma$ for $R$ small enough such that $\Sigma$ is spacelike. One can verify that
\begin{equation}
K(q) = 2a(2-d) \;\;\; , \;\;\; K_{11}(q) = 0 \;\;\; ,
\end{equation}
where we have evaluated the scalars at $q$, and we have taken the normal vector to be past-pointing, since this is a order reversed setup.

In the $Y^{\alpha}=0$ plane the surface $\Sigma$ is given by the $X^0=0$ line, and the extrinsic curvature scalar, $K$, is constant along this line (its value being $2a(2-d)$). The future-directed geodesics normal to $\Sigma$ within this plane are given by lines of constant $X^1$, and the proper time along these geodesics is simply $X^0$. The volume $V(X)$  can be written as a function of $X^0$ only:
\begin{align}\label{eq:cone_volume_formula}
V(X^0) & = \tilde{V}(X^0) \left(1 + \frac{d}{2(d+1)}K(q) X^0 + O(\varepsilon^2) \right) \nonumber
\\
& =\tilde{V}(X^0) \left(1 + \frac{d \, a(2-d)}{(d+1)} X^0 + O(\varepsilon^2) \right) \;\;\; 
\end{align}
using the cone volume formula in~\cite{Buck:2015oaa,Jubb:2016mbi}, and the flat cone volume $\tilde{V}(X^0)$ given in~\eqref{eq:flat_cone_volume}. The formula for $V(X^0)$ above is a special case of~\eqref{eq:expansion_of_volumes_in_paramters} in which $K(q)$ is the only non-zero scalar in the set $\lbrace \mathcal{G}_i(q)\rbrace$. To be explicit, let us set $\mathcal{G}_1(q) = K(q)$. We can  use the above expression for $V(X)$ to determine the form of the function  $f_1(X)$ that multiplies $\mathcal{G}_1(q)$ in~\eqref{eq:expansion_of_volumes_in_paramters}:
\begin{equation}
f_1(X) = \frac{d}{2(d+1)}X^0 \;\;\; .
\end{equation}
We can write down the volume integral for $V_+(X)$ in dimensions $d\geq 3$. It will be useful to first define $X_*^A$, as the $X^A$ coordinates at which the three surfaces ($\mathcal{P}$, $\mathcal{H}$ and $\Sigma$) meet. One can verify that
\begin{equation}
X_*^0 = \frac{a (X^0-X^1) (X^0+X^1)}{2 a (X^0+X^1)+1} \;\;\; , \;\;\; X_*^1 = -X_*^0  \;\;\; .
\end{equation}
For $X^1 < X_*^1$ the volume integral is
\begin{align}
V_+(X) & = \tilde{V}_+(X) \nonumber
\\
& - \int_{\mathbb{S}^{d-3}} d \Omega_{d-3} \int_0^{X_*^0} d {X'}^0 \int_{X_{\mathcal{P}\cap\Sigma,-}^1}^{-{X'}^0} d {X'}^1 \int_{\sqrt{\frac{{X'}^0}{a}}}^{\sqrt{(X^0-{X'}^0)^2-({X'}^1-X^1)^2}} dR^{d-3} \;\;\; ,
\end{align}
where $\tilde{V}_+(X)$ is given in~\eqref{eq:vplus_dfour}. Note that the limits of the $R$ integral are only defined for $a>0$, which is what we had assumed above. We have also introduced the notation $X_{\mathcal{P}\cap\Sigma,\pm}^1$ to denote the two values of ${X'}^1$ at which the surface $\mathcal{P}$ intersects $\Sigma$ (at a fixed value of ${X'}^0$). We have 
\begin{equation}
X_{\mathcal{P}\cap\Sigma,\pm}^1 = \pm \frac{\sqrt{-2 a {X'}^0 X^0+{X'}^0 (a {X'}^0-1)+a(X^0)^2}}{\sqrt{a}}+X^1 \;\;\; .
\end{equation}
For $X^1 \geq X_*^1 $ the volume integral is
\begin{align}
V_+(X) & = \tilde{V}_+(X) \nonumber
\\
& - \int_{\mathbb{S}^{d-3}} d \Omega_{d-3} \int_0^{X_{\mathcal{P}\cap\Sigma,*}^0} d {X'}^0 \int_{X_{\mathcal{P}\cap\Sigma,-}^1}^{X_{\mathcal{P}\cap\Sigma,+}^1} d {X'}^1 \int_{\sqrt{\frac{{X'}^0}{a}}}^{\sqrt{(X^0-{X'}^0)^2-({X'}^1-X^1)^2}} dR^{d-3} \nonumber
\\
& + \int_{\mathbb{S}^{d-3}} d \Omega_{d-3} \int_0^{X_*^0} d {X'}^0 \int_{-{X'}^0}^{X_{\mathcal{P}\cap\Sigma,+}^1} d {X'}^1 \int_{\sqrt{\frac{{X'}^0}{a}}}^{\sqrt{(X^0-{X'}^0)^2-({X'}^1-X^1)^2}} dR^{d-3} \;\;\; .
\end{align}
The result, in $d=4$, for both cases ($X^1 < X_*^1 $ and $X^1 \geq X_*^1 $) is
\begin{equation}
V_+(X) = \tilde{V}_+(X)\left(1 -\frac{8 a \left(8 (X^0)^2+9 X^0 X^1+3 (X^1)^2\right)}{5 (5 X^0+3 X^1)} + O(\varepsilon^2) \right) \;\;\; ,
\end{equation}
Similarly to $V(X)$, we can use this expression for $V_+(X)$ to determine the form of the function $f_{+,1}(X)$ in~\eqref{eq:expansion_of_volumes_in_paramters}. The resulting integrals can  be evaluated to determine the constant $b^{(4)}_{1,n}$:
\begin{align}
b^{(4)}_{1,n} = &  -\frac{3^{-3 n-\frac{5}{4}} 4^{-2 n}}{5 \pi ^{3/4} n! \Gamma (4 n+2)}\Bigg\lbrace \frac{3\ 4^{6 n+2} \Gamma \left(n+\frac{11}{4}\right)}{7+4 n} \Bigg[
n! (3 n)! \nonumber
\\
& - \Gamma (4 n+2) \Big( -B_{\frac{3}{4}}(3 n+1,n+1)+B_{\frac{3}{8}}(3 n+1,n+1) \nonumber
\\
& +B_{\frac{5}{8}}(n+1,3 n+1) \Big)
 \Bigg]  - \frac{\Gamma \left(n+\frac{7}{4}\right)}{(4 n+1) (4 n+3)} \Bigg[ 8^{4 n+1} (4 n+1) (12 n+1) n! (3 n)! \nonumber
\\
& + n \Gamma (4 n+2) \Big( 
8^{4 n+1} \left((12 n-1) B_{\frac{1}{4}}(n,3 n+1)-18 n B_{\frac{1}{4}}(n,3 n)\right) \nonumber
\\
& -3^{3 n+2} 5^n (4 n+1) \big[ F_1(2;-3 n,-n;3;-1,3/5) \nonumber
\\
& -2 (3 n)! \, _2\tilde{F}_1 (2,-n;3 n+3;-3/5 )\big]
\Big) \Bigg] \Bigg\rbrace \;\;\; ,
\end{align}
where $F_1(a; b_1 , b_2 ; c ; x,y)$ is the Appell hypergeometric function of two variables, and where $_2\tilde{F}_1(a, b; c;z)$ is the regularised hypergeometric function. In terms of the hypergeometric function $_2F_1(a, b; c;z)$, one defines the regularised function as $_2\tilde{F}_1(a, b; c;z):=_2\hspace{-1mm}F_1(a, b; c;z)/\Gamma(c)$.

This complicated expression greatly simplifies when one considers specific values of $n$. For example,
\begin{align}
b^{(4)}_{1,1} & = -\frac{4}{175} \left(\frac{3}{\pi }\right)^{3/4} \Gamma \left(\frac{11}{4}\right) \approx -0.0355127\;\;\; ,
\\
b^{(4)}_{1,2} & = -\frac{26 \, \Gamma \left(\frac{15}{4}\right)}{1925 \sqrt[4]{3} \pi ^{3/4}} \approx -0.019236 \;\;\; ,
\\
b^{(4)}_{1,3} & = -\frac{31 \left(\frac{3}{\pi }\right)^{3/4} \Gamma \left(\frac{15}{4}\right)}{10010} \approx -0.0132319 \;\;\; .
\end{align}

\subsubsection{$b^{(d)}_{2,n}$}

Here we take the same setup as above, but with $\Sigma$ given by the zeroes of the function
\begin{equation}
S_{\Sigma}(Z)=X^0-a (X^1)^2 \;\;\;
\end{equation}
for $a>0$. Note that this equation only applies for $X^1$ small enough such that $\Sigma$ is spacelike. The only non-zero components of the past-pointing normal vector $n$ are 
\begin{equation}
n^0 = -\frac{1}{\sqrt{1-4a^2(X^1)^2}} \;\;\; ,
\;\;\;
n^1 = -\frac{2a X^1}{\sqrt{1-4a^2(X^1)^2}} \;\;\; ,
\end{equation}
where these vectors live in the tangent space of a point $Z^a=(X^0,X^1,Y^{\alpha})=(a(X^1)^2,X^1,0)$, i.e. a point on $\Sigma$. The resulting scalars $K$ and $K_{11}$, at any point on $\Sigma$, are
\begin{equation}
K = K_{11} = -2a(1-4a^2(X^1)^2)^{-\frac{3}{2}} \;\;\; , 
\end{equation}
and hence their values at $q$, i.e. the origin, are
\begin{equation}
K(q) = K_{11}(q) = -2a \;\;\; .
\end{equation}

The normal geodesics from $\Sigma$ within the $Y^{\alpha}$ plane will remain within the plane. They will be straight lines of the form
\begin{equation}\label{eq:normal_geodesic}
Z^a(\tau) = Z_0^a - n^a(Z_0) \tau \;\;\; ,
\end{equation}
where we have written the normal vector as a function of the point $q_0$ (with coordinates ${Z_0}^a=({X_0}^0,{X_1}^1,0) = (a({X_0}^1)^2,{X_1}^1,0)$) at which the geodesic intersects $\Sigma$. The minus sign is there so that $\tau$ is the proper time to the future of $\Sigma$ (the normal $n$ is past-pointing).

The function $I_n^{(d)}(q; l, \tau)$ that we wish to evaluate involves integrals over the $X^A$ coordinates of a point in the plane normal to $\mathcal{J}$. In order to evaluate these integrals we need to express the cone volume, $V(X)$, in terms of the coordinates $X^A$ of a point $q_1$ in this plane. Let $\tau$ be the proper time along a normal geodesic (of the form~\eqref{eq:normal_geodesic}) that intersects the point $q_1$, and starts at $q_0$ on $\Sigma$. From~\cite{Buck:2015oaa} we know that the cone volume can be expressed in terms of $\tau$ as
\begin{equation}
V(\tau) = \tilde{V}(\tau) \left(1 + \frac{d}{2(d+1)}K(q_0) \tau + O(\varepsilon^2) \right) \;\;\; ,
\end{equation}
where we have evaluated $K$ at $q_0$. We can write $K$ as a function of the coordinates, ${X_0}^A$, of $q_0$ as
\begin{equation}
K(X_0) =  -2a(1-4a^2({X_0}^1)^2)^{-\frac{3}{2}} \;\;\; , 
\end{equation}
using the above expression for $K$ at any point on $\Sigma$.

In order to rewrite $V(\tau)$ as a function of $X^A$ we need to solve for $\tau$ and ${X_0}^1$ in terms of the coordinates $X^A$. Explicitly, we have to solve the following equations:
\begin{align}
X^0 & = a ({X_0}^1)^2 +\frac{\tau}{\sqrt{1-4a^2({X_0}^1)^2}} \;\;\; , \nonumber
\\
X^1 & = {X_0}^1 + \frac{2a {X_0}^1 \tau}{\sqrt{1-4a^2({X_0}^1)^2}} \;\;\; ,
\end{align}
for $\tau$ and ${X_0}^1$. To the relevant order, one finds
\begin{align}\label{eq:tau_xq_in_terms_of_tp_xP}
\tau & = X^0 - a (X^1)^2 + O(\varepsilon^2) \;\;\; ,
\\
{X_0}^1 & = X^1 - 2 a  X^1 \tau + O(\varepsilon^2) \;\;\; .
\end{align}
We can  write $K$ as a function of $X^A$:
\begin{equation}\label{eq:kScalar_in_terms_of_xp}
K(X) = -2a -12a^3 (X^1)^2 + O(\varepsilon^3) \;\;\; .
\end{equation}
The cone volume can  be expressed as a function of $X^A$, using equations~\eqref{eq:tau_xq_in_terms_of_tp_xP} and~\eqref{eq:kScalar_in_terms_of_xp}. In $d=2$ and $4$ we have
\begin{align}
d=2 \;\;\; : \;\;\; V(X) & = \tilde{V}(X^0) \left(1 -\frac{2 a \left((X^0)^2+3 (X^1)^2\right)}{3 X^0} + O(\varepsilon^2)\right) \;\;\; ,
\\
d=4 \;\;\; : \;\;\; V(X) & = \tilde{V}(X^0) \left(1 -\frac{4 a \left((X^0)^2+5 (X^1)^2\right)}{5 X^0} +O(\varepsilon^2) \right) \;\;\; .
\end{align}
We can also determine the volume $V_+(X)$, in  $d=2$ and $4$. In both cases it will be useful to introduce $X^A_{\mathcal{P}\cap\Sigma}$ as the smallest $X^1$ value at which $\mathcal{P}$ intersects $\Sigma$. Explicitly,
\begin{equation}
X^1_{\mathcal{P}\cap\Sigma} = -\frac{\sqrt{4 a (X^0-X^1)+1}-1}{2 a} \;\;\; .
\end{equation}
In $d=2$ the volume integral can  be written as
\begin{align}
V_+(X) & = \int_{X^1_{\mathcal{P}\cap\Sigma}}^{\frac{1}{2}(X^1-X^0)} d{X'}^1 \int_{a ({X'}^1)^2}^{{X'}^1+X^0-X^1} d{X'}^0 \nonumber
\\
& + \int_{\frac{1}{2}(X^1-X^0)}^{0} d{X'}^1 \int_{a ({X'}^1)^2}^{-{X'}^1} d{X'}^0 \;\;\; ,
\end{align}
and for $d\geq 3$ we have
\begin{align}
V_+(X) & = \int_{\mathbb{S}^{d-3}} d \Omega_{d-3} 
\int_{0}^{a (X^1_{\mathcal{P}\cap\Sigma})^2}d{X'}^0 
\int_{-\sqrt{\frac{{X'}^0}{a}}}^{-{X'}^0} d{X'}^1
\int_{0}^{\sqrt{(X^0-{X'}^0)^2-({X'}^1-X^1)^2}} dR^{d-3}
\nonumber
\\
& +\int_{\mathbb{S}^{d-3}} d \Omega_{d-3} 
\int_{a (X^1_{\mathcal{P}\cap\Sigma})^2}^{\frac{1}{2}(X^0-X^1)}d{X'}^0 
\int_{X^1 - X^0 + {X'}^0}^{-{X'}^0} d{X'}^1
\int_{0}^{\sqrt{(X^0-{X'}^0)^2-({X'}^1-X^1)^2}} dR^{d-3} \;\;\; .
\end{align}
Evaluating these integrals in $d=2$ and $d=4$ we get
\begin{align}
d=2 \;\;\; : \;\;\; V_+(X) & = \tilde{V}_+(X) \left(1 -\frac{4}{3} a (X^0-X^1) + O(\varepsilon^2)\right) \;\;\; ,
\\
d=4 \;\;\; : \;\;\; V_+(X) & = \tilde{V}_+(X) \left(1 -\frac{8 a (X^0-X^1) (4 X^0+X^1)}{5 (5 X^0+3 X^1)} +O(\varepsilon^2) \right) \;\;\; .
\end{align}

We  have everything we need to evaluate the first order correction to $I_n^{(d)}(q; l, \tau)$. In $d=2$ we must match the first order correction to a term of the form
\begin{equation}
b^{(2)}_{n}K(q) \, l \;\;\; ,
\end{equation}
in order to determine the constant $b^{(2)}_{n}$ (we must also use the fact that $K(q)=-2a$). We find
\begin{equation}
b^{(2)}_{n} = -\frac{2 \Gamma \left(n+\frac{5}{2}\right)}{6 n \Gamma (n+2)+3 \Gamma (n+2)} \;\;\; .
\end{equation}
In $d=4$ we must match the first order correction to a term of the form
\begin{equation}
\left( b^{(4)}_{1,n}K(q) + b^{(4)}_{2,n}K_{11}(q) \, \right)l \;\;\; ,
\end{equation}
and we must use our existing expression for $b^{(4)}_{1,n}$ to solve for the constant $b^{(4)}_{1,n}$ (we must also use the fact that $K(q)=K_{11}(q)=-2a$). The resulting expression is even longer than the expression for $b^{(4)}_{1,n}$, so we will not write it here. Instead, we will give the much simplified expressions one gets for specific values of $n$:
\begin{align}
b^{(4)}_{2,1} & = -\frac{2}{35} \left(\frac{3}{\pi }\right)^{3/4} \Gamma \left(\frac{11}{4}\right) \approx -0.0887817\;\;\; ,
\\
b^{(4)}_{2,2} & = -\frac{1}{77} \left(\frac{3}{\pi }\right)^{3/4} \Gamma \left(\frac{15}{4}\right) \approx -0.0554886 \;\;\; ,
\\
b^{(4)}_{2,3} & = -\frac{83 \Gamma \left(\frac{19}{4}\right)}{10725 \sqrt[4]{3} \pi ^{3/4}} \approx -0.0413319 \;\;\; .
\end{align}

\subsubsection{$b^{(d)}_{3,n}$}

In this section we use the same setup as above, but we will take $\Sigma$ to be the surface given by $X^0=0$, so that $K=K_{11}=0$. To get a non-zero null expansion, $\theta$, we take $\mathcal{H}$ to be the past lightcone of a point with coordinates $Z^a=(X^0,X^1,Y^{\alpha})=(r,-r,0)$, for $r>0$. This past lightcone will pass through the point $q$ at the origin. We can describe $\mathcal{H}$ by the equation
\begin{equation}
X^0=r-\sqrt{(r+X^1)^2+R^2} \;\;\; ,
\end{equation}
where $R$ is the radius in the $Y^{\alpha}$ directions introduced above. Using~\eqref{eq:theta_definition} we find that
\begin{equation}
\theta(q) = \frac{d-2}{\sqrt{2}\,r} \;\;\; .
\end{equation}

In this setup the volume $V(X)$ is simply the flat volume $\tilde{V}(X^0)$. For $d\geq 3$ we can write down the volume integral for $V_+(X)$ as
\begin{align}
V_+(X) & = \tilde{V}_+(X) \nonumber
\\
& -  \int_{\mathbb{S}^{d-3}} d \Omega_{d-3} \int_0^{\frac{1}{2}(X^0-X^1)} d{X'}^0 \int_{X^1_{\mathcal{P}\cap\mathcal{H}}}^{-{X'}^0} d{X'}^1 \int_{R_{\mathcal{H}}}^{R_{\mathcal{P}}} dR^{d-3} \;\;\; ,
\end{align}
where
\begin{equation}
X^1_{\mathcal{P}\cap\mathcal{H}} = \frac{-2 r {X'}^0+2 {X'}^0 X^0-(X^0)^2+(X^1)^2}{2 (r+X^1)} \;\;\; ,
\end{equation}
is the $X^1$ value at which $\mathcal{P}$ intersects $\mathcal{H}$, for a fixed value of ${X'}^0$, and where we have defined
\begin{align}
R_{\mathcal{H}} & : = \sqrt{({X'}^0- 2 r -  {X'}^1) ({X'}^0 + {X'}^1)} \;\;\; ,
\\
R_{\mathcal{P}} & : = \sqrt{({X'}^0 - X^0 + X^1 - {X'}^1) ({X'}^0 -X^0 - X^1 + {X'}^1)} \;\;\; .
\end{align}
Evaluating this volume integral in $d=4$ gives
\begin{equation}
V_+(X) = \tilde{V}_+(X)\left(1 -\frac{2 (X^0+X^1)^2}{r (5 X^0+3 X^1)} + O(\varepsilon^2) \right) \;\;\; .
\end{equation}
Following similar steps to above, we can evaluate the first order correction to $I_n^{(d)}(q; l, \tau)$, and match it to an expression of the form
\begin{equation}
b^{(4)}_{3,n}\theta(q) \, l \;\;\; ,
\end{equation}
to determine the constant $b^{(4)}_{3,n}$. We find
\begin{align}
b^{(4)}_{3,n} & = \frac{3^{-3 n-\frac{9}{4}} 4^{-2 n} \pi ^{-n-\frac{3}{4}} \Gamma \left(n+\frac{7}{4}\right)}{\sqrt{2} (4 n+1) (4 n+3) n! \Gamma (4 n+2)} \Bigg\lbrace 
4^{6 n+1} \pi ^n \Bigg[ -2 (4 n+1) (6 n+1) n! (3 n)! \nonumber
\\
& +n\Gamma (4 n+2) \Big(9 n B_{\frac{1}{4}}(n,3 n)+2 B_{\frac{1}{4}}(n,3 n+1) \Big) \Bigg] \nonumber
\\
& + 3^{3 n+2} n (4 n+1) (5 \pi )^n \Gamma (4 n+2) \Bigg[ F_1\left(2;-3 n,-n;3;-1,\frac{3}{5}\right)
\\
& -2 (3 n)! \, _2\tilde{F}_1\left(2,-n;3 n+3;-\frac{3}{5}\right) \Bigg]
\Bigg\rbrace \;\;\; .
\end{align}
For particular values of $n$ we get
\begin{align}
b^{(4)}_{3,1} & =-\frac{\sqrt{2}\, \Gamma \left(\frac{11}{4}\right)}{35 \sqrt[4]{3} \pi ^{3/4}} \approx -0.0209261\;\;\; ,
\\
b^{(4)}_{3,2} & = -\frac{19 \, \Gamma \left(\frac{15}{4}\right)}{1155 \sqrt{2} \sqrt[4]{3} \pi ^{3/4}} \approx -0.0165665 \;\;\; ,
\\
b^{(4)}_{3,3} & = -\frac{271 \, \Gamma \left(\frac{19}{4}\right)}{75075 \sqrt{2} \sqrt[4]{3} \pi ^{3/4}} \approx -0.0136321 \;\;\; .
\end{align}

\section{Causal set geometry}

\subsection{Extracting the horizon area}

We have  determined that $I_n^{(d)}(q; l, \tau)$ has the small $l$ expansion given in~\eqref{eq:integral_expansion_claim} for $d>2$, and~\eqref{eq:integral_expansion_claim_2d} for $d=2$. For $d=2$ and $d=4$ we have determined explicit expressions for the coefficients $a_n^{(d)}$ (and $a_1^{(3)}$), and we have determined the constants $b^{(2)}_n$ and $b^{(4)}_{i,n}$ that appear at first order in $l$. In this section we will discuss how to use the explicit expressions for these constants to extract continuum geometry from the causal set.

The simplest geometrical quantity to extract is the horizon area. If we are given a causal set, $\mathcal{C}$, and the corresponding partitions $\mathcal{C}^{\pm}_{\pm}$,  we can count the number of horizon molecules $\mathrm{H}$ and calculate
\begin{equation}
\frac{\rho^{\frac{2-d}{d}}}{a^{(d)}} \mathrm{H} \;\;\; ,
\end{equation}
using our above expressions for $a^{(d)}=a^{(d)}_1$ in $d=2,3$, and $4$. If this causal set has come from a sprinkling into a spacetime with a horizon, then this value corresponds to the causal set estimate of the continuum horizon area. Under the sprinkling process, this value, $\frac{\rho^{\frac{2-d}{d}}}{a^{(d)}} \mathrm{H}$, becomes the random variable $\frac{\rho^{\frac{2-d}{d}}}{a^{(d)}} \mathbf{H}$, and from our above arguments we know its expectation value has the following limit 
\begin{equation}
\lim_{\rho\rightarrow \infty} \left\langle\frac{\rho^{\frac{2-d}{d}}}{a^{(d)}} \mathbf{H} \right\rangle = \int_{\mathcal{J}} dV_{\mathcal{J}} \;\;\; .
\end{equation}
That is, it gives the horizon area in the continuum limit. Two questions remain: \textit{i)} is the value $\frac{\rho^{\frac{2-d}{d}}}{a^{(d)}} \mathrm{H}$, for a single causal set, close to the continuum horizon area, and \textit{ii)} for a finite, but small, $l$ relative to the curvature scales of the setup, is the expectation value $\left\langle\frac{\rho^{\frac{2-d}{d}}}{a^{(d)}} \mathbf{H} \right\rangle$ close to the continuum horizon area?

The second question can be answered, to some extent, immediately, as we have determined the first order correction to $I_n^{(d)}(q; l, \tau)$. Recall that we can write the expectation value of $\mathbf{H}$ in terms of $I_n^{(d)}(q; l, \tau)$ as (using~\eqref{eq:expectation_in_term_of_I})
\begin{equation}
\rho^{\frac{2-d}{d}} \langle \mathbf{H} \rangle = \int_{\mathcal{J}}dV_{\mathcal{J}} \, I_1^{(d)}(q; l, \tau) + \cdots \;\;\; ,
\end{equation}
where ``$\cdots$" denote exponentially suppressed terms in $\rho$. We have also written this integral in a more geometric way than~\eqref{eq:expectation_in_term_of_I}, without referring to any particular coordinates on $\mathcal{J}$. We can  use the small $l$ expansion of $I_1^{(d)}(q; l, \tau)$ to see that
\begin{equation}
 \left\langle\frac{\rho^{\frac{2-d}{d}}}{a^{(d)}} \mathbf{H} \right\rangle = \int_{\mathcal{J}} dV_{\mathcal{J}} + \frac{l}{a^{(d)}} \int_{\mathcal{J}} dV_{\mathcal{J}} \left( b^{(d)}_{1,1}K+b^{(d)}_{2,1}K_{11} + b^{(d)}_{3,1}\theta \right) + O(l^2) \;\;\; ,
\end{equation}
where the scalars $K$, $K_{11}$, and $\theta$, depend on the point $q\in\mathcal{J}$, and so they may vary across the integral. The expectation value on the left will  be close to the continuum horizon area if the first order correction is small, that is, if
\begin{equation}
\frac{l}{a^{(d)}} \int_{\mathcal{J}} dV_{\mathcal{J}} \left( b^{(d)}_{1,1}K+b^{(d)}_{2,1}K_{11} + b^{(d)}_{3,1}\theta \right) <<   \int_{\mathcal{J}} dV_{\mathcal{J}}  \;\;\; .
\end{equation}
This will be satisfied if $l$ is much less than any of the curvature scales of the setup, for all points $q\in\mathcal{J}$.

The first question above is more difficult, as it requires us to look at how the random variable $\mathbf{H}$ fluctuates under the sprinkling process. If the fluctuations are large, then the value $\frac{\rho^{\frac{2-d}{d}}}{a^{(d)}} \mathrm{H}$, for a single causal set, will likely be very different from the continuum horizon area. One may be able to estimate the fluctuations numerically in specific geometrical setups. We have not attempted such an investigation here, and so we leave the first question as an open problem for future work.

\subsection{Extracting other geometry}

In the last section we found that we could count $\frac{\rho^{\frac{2-d}{d}}}{a^{(d)}} \mathrm{H}$ to get an estimate for the horizon area of a causal set. In the continuum limit the expectation value of the associated random variable was the horizon area, which is proportional to the first term in the small $l$ expansion of
\begin{equation}
\int_{\mathcal{J}}dV_{\mathcal{J}} \, I_1^{(d)}(q; l, \tau)
=
a^{(d)}_n \int_{\mathcal{J}}dV_{\mathcal{J}} + l \int_{\mathcal{J}} dV_{\mathcal{J}} \left( b^{(d)}_{1,1}K+b^{(d)}_{2,1}K_{11} + b^{(d)}_{3,1}\theta \right) + O(l^2)  \;\;\; .
\end{equation}
We can  ask if it is possible to extract the second term in this expansion (the term of $O(l)$) using the causal set. That is, can we extract the geometrical quantity
\begin{equation}
\int_{\mathcal{J}} dV_{\mathcal{J}} \left( b^{(d)}_{1,1}K+b^{(d)}_{2,1}K_{11} + b^{(d)}_{3,1}\theta \right) \;\;\; ,
\end{equation}
by counting something on the causal set.

Following the procedure given in~\cite{Buck:2015oaa,Jubb:2016mbi}, we can get close to extracting the first order correction using the following causal set random variable:
\begin{equation}\label{eq:higher_order_geometry_random_variable}
l^{d-3}\left( \frac{\mathbf{H_n}}{a^{(d)}_n}  - \frac{\mathbf{H_m}}{a^{(d)}_m} \right) \;\;\; ,
\end{equation}
where $n \neq m$. From the expansion for $I_n^{(d)}(q; l, \tau)$ we can determine the expectation value of this random variable. We find
\begin{equation}
\left\langle l^{d-3}\left( \frac{\mathbf{H_n}}{a^{(d)}_n}  - \frac{\mathbf{H_m}}{a^{(d)}_m} \right) \right\rangle = \int_{\mathcal{J}} dV_{\mathcal{J}} \left( b^{(d)}_{1,nm}K+b^{(d)}_{2,nm}K_{11} + b^{(d)}_{3,nm}\theta \right) + O(l) \;\;\; ,
\end{equation}
where
\begin{equation}
b^{(d)}_{i,nm} : = \frac{b^{(d)}_{i,n}}{a^{(d)}_n} - \frac{b^{(d)}_{i,m}}{a^{(d)}_m} \;\;\; .
\end{equation}
The random variable in~\eqref{eq:higher_order_geometry_random_variable} is not as obviously useful as $\frac{\rho^{\frac{2-d}{d}}}{a^{(d)}} \mathbf{H}$, but it may be more useful in the future when combined with other causal set expressions for extracting continuum geometry. Perhaps the most interesting thing to note from this expression is that, for the first time, a causal set expression has been found that depends upon the null expansion, $\theta$, of some null surface.

\section{Entropy}

Dou and Sorkin suggest that horizon molecule identification and counting in a causal set bears the same relation to the black hole entropy as does the counting of molecules of a gas to the entropy of the gas. The fact that we get the right dependence on the area and the right order of magnitude, if the discreteness length is of order the Planck length, is encouraging. We will not know whether our molecules are the ``right'' ones, however, until we know the statistical mechanics of black hole thermodynamics within the full theory of quantum causal sets, in which the entropy is understood in terms of the number of microstates corresponding to the macrostate of the black hole. 

Other plausible molecule definitions are indeed possible to find. Some involve the causet to the future of $\Sigma$. For example, we can take as a horizon molecule a link $p\prec q$ in which $p$ is in $\mM_-^-$ and is maximal in the past of $\Sigma$, and $q$ is in $\mM_+^+$ and is minimal in the future of $\Sigma$.  Similar locality arguments to those we have made in this paper can be made for the claim that the expected number of these molecules will also give the area of $\mJ$ in discreteness units, up to a (different) factor of order one. 
It may be that when we fully understand black hole entropy it will pick out one molecule definition, or it may 
turn out that no one definition of horizon molecule is favoured over any other that works at this level. 

The most promising aspect of our investigation is that the result is universal for all causal horizons in any particular dimension.  Following  Jacobson and Parentani, it supports the idea that the thermodynamics of black holes is just one aspect of the thermodynamics of causal horizons in general. The result reported here is an encouragement to look for a universal statistical mechanics of causal horizons, so that black hole entropy, cosmological horizon entropy, and Rindler horizon entropy, all find a unified explanation.

\section*{Acknowledgements}

We thank Jeremy Butterfield for useful discussions. This research was supported in part by Perimeter Institute for Theoretical Physics. Research at Perimeter Institute is supported by the Government of Canada through Industry Canada and by the Province of Ontario through the Ministry of Economic Development and Innovation. FD is supported in part by STFC grant ST/P000762/1 and APEX grant APX/R1/180098. IJ is supported by an Irish Research Council Fellowship (GOIPD/2018/180).

\begin{appendices}

\section{$I^-(\mathcal{M}^-_+)\cap \mathcal{M}^-_-=I^-(\mathcal{J})$}\label{app:proof_mj_is_past_of_j}

\begin{proof}
\begin{itemize}
\item[\phantom{a}]

\item[(:$\supseteq$:)]

For any point $p\in I^-(\mathcal{J})$, there exists a point $q\in\mathcal{J}$ such that $p\ll q$ (the notation means $p$ is to the chronological past of $q$). So,  $q\in I^+(p)$, and as $I^+(p)$ is open, there exists an open neighbourhood $\mathcal{O}$ of $q$ such that $\mathcal{O}\subset I^+(p)$. As $\mathcal{J}$ lies on the boundary of $\mathcal{M}^-_+$ and of $\mathcal{M}^-_-$, we have $\mathcal{M}^-_+ \cap \mathcal{O} \neq \emptyset$ and $\mathcal{M}^-_- \cap \mathcal{O} \neq \emptyset$. Therefore, $p\in I^-(\mathcal{M}^-_{\pm})$. Now,  $I^-(\mathcal{M}^-_-) =\mathcal{M}^-_-$, and so $p\in \mathcal{M}^-_-$.

\item[(:$\subseteq$:)]

For any point $p\in I^-(\mathcal{M}^-_+)\cap \mathcal{M}^-_- $ there exists a future directed timelike curve $\gamma$ from $p\in\mathcal{M}^-_-$ to some point $p'\in \mathcal{M}^-_+$. Such a curve must pass through $\mH$ at a single point $q\in \mH$ (proposition 3.15.~\cite{Penrose}), and $p \ll q \ll p'$, so $q$ is to the past of
$\Sigma$. $q$ lies on a future inextendible null geodesic generator of $\mathcal{H}$, which must pass though a point $q'$ in $\Sigma$.
So $q' \in \mH \cap \Sigma = \mJ$.
 As $p \ll q$, we have  $p\ll q'$, and so $p\in I^-(\mJ)$.
\end{itemize}
\end{proof}

\section{Determining the set of independent scalars}\label{app:determining_the_set_of_scalars}

To systematically find all the independent scalars of dimension $L^{-1}$, we must consider all possible first order derivatives of contractions of tensors that depend upon the basic dimensionless geometrical objects of our setup. The basic dimensionless geometrical objects are the metric $g$, the future-pointing normal vector $n$ on $\Sigma$ (normalised as $n.n=-1$), and the future-pointing null vector $k=d/d\lambda$ on $\mathcal{H}$ (where the affine parameter $\lambda$ is chosen such that $k.n=-1/\sqrt{2}$ on $\mathcal{J}$). We also have the spacelike vector $m=\sqrt{2}k - n $ on $\mathcal{J}$, which is tangent to $\Sigma$ and orthogonal to $\mathcal{J}$. Note that $m.m=1$ and $n.m=0$. Lastly, we have the null vector $l$ on $\mathcal{H}$ such that $l.k=-1$, and such that $l$ is orthogonal to all the coordinate vectors $\partial/\partial y^{\alpha}$, where $y^{\alpha}$ are the FSNC's defined above. Let us denote the above set of tensors by $\mathfrak{G}:=\lbrace g, n, m, k, l \rbrace$.

It is worth noting that we cannot consider any tensors that are independent of those in $\mathfrak{G}$. Such a tensor, by definition, would be unchanged as the tensors in $\mathfrak{G}$ vary. This means that this tensor would be constant under changes to the spacetime geometry, and the embeddings of the sub-manifolds $\Sigma$ and $\mathcal{H}$. That is, it would be entirely independent of our geometric setup, and hence any geometric quantity (such as the volumes $V(X)$ and $V_+(X)$) will be independent from it. An example of such a tensor would be an arbitrarily chosen vector lying within the tangent space of $\mathcal{J}$.

Any tensor that depends upon $\mathfrak{G}$ must also be some linear combination of tensor products and/or contractions of tensors in $\mathfrak{G}$ (it cannot be anything else if it is to be a tensor itself). Let us call this space of tensors $\mathfrak{A}$. To get the right dimensions of length we  consider covariant first order derivatives of tensors $A\in\mathfrak{A}$. The product rule  reduces the derivative of a given $A\in\mathfrak{A}$ to a linear combination of tensors that are each of the form of a single derivative of one of the tensors in $\mathfrak{G}$, contracted with, or in a tensor product with, some other $A''\in\mathfrak{A}$. Lastly, to form a scalar, we must  contract any remaining indices with some other tensor $A'''\in\mathfrak{A}$. Any index contractions that do not involve the index of the covariant derivative, and do not involve the index of the tensor inside the covariant derivative, will simply result in some constant. Therefore, the space of possible scalars consists of linear combinations of tensors formed from a single covariant derivative of one of the tensors in $\mathfrak{G}$, contracted with the minimum number of tensors in $\mathfrak{G}$ needed to form a scalar. Recall that we also wish to evaluate the resulting scalars at some point $q\in\mathcal{J}$.

Scalars of length dimension $L^{-1}$ will not involve first order derivatives of $g$, as its covariant derivative vanishes. Therefore, we can focus on covariant derivatives of $n$, $k$, $l$, and $m$. In components, these first order derivatives look like $n_{a;b}$, $k_{a;b}$, $l_{a;b}$, and $m_{a;b}$. We  need to form scalars from these four tensors using contractions with the minimum number of tensors from $\mathfrak{G}$. Before doing this, it should be noted that these covariant derivatives are not technically well-defined, as they involve derivatives of $n$, $m$, $k$, and $l$, in directions away from the surfaces on which they are defined. Therefore, we must project the derivatives onto the relevant surfaces using $h^a_{\; b}$, $\sigma^a_{\; b}$, $m^a$, and $k^a$. As $n$ is only defined on $\Sigma$, we must project the derivative onto the tangent space of $\Sigma$. This can be done in the following three ways:
\begin{equation}
n_{a;b} m^b \;\;\; , \;\;\; n_{a;b}h^b_{\; c} \;\;\; , \;\;\; n_{a;b}\sigma^b_{\; c} \;\;\; .
\end{equation}
$k$ and $l$ are defined on $\mathcal{H}$, and so we must project their derivatives onto the tangent space of $\mathcal{H}$ with $\sigma^a_{\; b}$ or $k^a$:
\begin{equation}
k_{a;b}k^b \;\;\; , \;\;\; k_{a;b}\sigma^b_{\; c} \;\;\; , \;\;\; l_{a;b}k^b \;\;\; , \;\;\; l_{a;b}\sigma^b_{\; c} \;\;\; .
\end{equation}
Lastly, $m$ is only defined on $\mathcal{J}$, so we must project the derivative onto the tangent space of $\mathcal{J}$ with $\sigma^b_{\; c}$, i.e. we have 
\begin{equation}
m_{a;b}\sigma^b_{\; c} \;\;\; .
\end{equation}
We  have 8 well-defined first order derivatives which we can contract with any of the tensors in $\mathfrak{G}$.

Starting with the covariant derivative of $n$ we have $n_{a;b}m^b$. To form a scalar we must contract with another vector. If we contract with $n^a$ we get $n^an_{a;b}m^b = (n.n)_{;b}m^b - n^an_{a;b}m^b = -n^an_{a;b}m^b$, and hence $n^an_{a;b}m^b=0$ (here we have used the fact that $n.n=-1$ on $\Sigma$). A contraction with $k^a$ will give the same result as a contraction with $m^a/\sqrt{2}$, since $k^a=1/\sqrt{2}(n^a+m^a)$, and since the contraction with $n^a$ vanishes. We can, therefore, focus on contracting $n_{a;b}m^b$ with $m^a$. The result is the component of the extrinsic curvature tensor in the $m$-direction, i.e. $K_{11}:=K_{ab}m^am^b$, where $K_{ab} = n_{(c;d)}h^c_{\; a} h^d_{\; b}$.

Next we have $n_{a;b}h^b_{\; c}$, which must be contracted with two upstairs indices. This can be done with two vectors, or with $g^{ab}$. The only vector we can use is $m^a$, as the $h^b_{\; c}$ in $n_{a;b}h^b_{\; c}$ will project $n$, $k$, and $l$ to some (possibly zero) multiple of $m^a$. If we contract with $m^a m^b$ then we will recover the component $K_{11}$ again, and so this is not an independent scalar. A contraction with $g^{ab}$ yields the extrinsic curvature scalar $K=h^{ab}K_{ab}$. 

The last expression involving a covariant derivative of $n$ is $n_{a;b}\sigma^b_{\; c}$. The $\sigma^b_{\; c}$ in this expression will kill any of the vectors $n$, $m$, $k$, and $l$, and hence we must contract the two free indices with $g^{ab}$. The result is $n_{a;b}\sigma^{ab}=K-K_{11}$ (one can verify this using the fact that $\sigma^a_{\; b} = h^a_{\; b} -m^a m_b$), and so it is not independent of the other scalars we have already mentioned.

Moving on to covariant derivatives of $k$, we have that $k_{a;b}k^b=0$, as the null curves ruling $\mathcal{H}$ are affinely parameterised geodesics. The next term to consider is $k_{a;b}\sigma^b_{\; c}$, which must be contracted with two upstairs indices. The $\sigma^b_{\; c}$ in this expression will kill any vector that we can contract with, and hence we must contract both the free indices with $g^{ab}$. The result is simply the null expansion $\theta = k_{a;b}\sigma^{ab}$.

The first expression to consider for $l$ is $l_{a;b}k^b$. As $k_{a;b}k^b=0$ we have that $k_{a;b}l^a k^b=0$, and hence that
\begin{equation}
0=k_{a;b}l^a k^b = (k.l)_{;b}k^b - l_{a;b}k^a k^b = - l_{a;b}k^a k^b  \;\;\; ,
\end{equation}
where we have used the fact that $k.l=-1$ on $\mathcal{H}$. Since $l_{a;b}k^a k^b =0 $, we know that $l_{a;b}k^b$ is a covector within the cotangent space of $\mathcal{H}$. The only vector within the tangent space of $\mathcal{H}$ that we can contract it with is $k^a$, but we have just seen that this contraction vanishes. Therefore, the term $l_{a;b}k^b$ will not give us any new independent scalars.

Next we have $l_{a;b}\sigma^b_{\; c}$. Using the fact that $l=\sqrt{2}n - k$, and the contractions already considered above, one can see that this will not give anything new. For the last term $m_{a;b}\sigma^b_{\; c}$, we can use the fact that $m=\sqrt{2}k-n$ to show that this will also give nothing new.

In summary, we can only form three independent scalars at a point $q\in\mathcal{J}$ involving a single derivative: $\theta$, $K_{11}$, and $K$.

\end{appendices}

\newpage
\printbibliography[heading=bibnumbered]

\end{document}